\documentclass[showpacs,amsmath,twocolumn,floatfix]{revtex4-1}

\usepackage{subfigure}
\usepackage{graphicx}
\usepackage{amssymb}
\usepackage{amsmath}
\usepackage{bbm}
\usepackage{stmaryrd}
\usepackage{rotating}
\DeclareMathAlphabet{\cali}{OMS}{zplm}{m}{n}

\newcommand\ii{\mathrm{i \,}}
\newcommand\dd{\delta\:\!}

\begin{document}

\title{Collective modes in the paramagnetic phase of the Hubbard model}
\author{Vu Hung Dao and Raymond Fr\'esard}
\affiliation{Normandie Univ, ENSICAEN, UNICAEN, CNRS, CRISMAT, 14000 Caen, France}
%\affiliation{Laboratoire CRISMAT, CNRS UMR6508, Normandie Universit\'e,
%6 Boulevard Mar\'echal Juin, 14050 Caen Cedex 4, France}
\keywords{Hubbard model, slave boson, collective mode}
\pacs{71.10.Fd, 72.15.Nj, 71.30.+h}

\begin{abstract}
The charge dynamical response function of the Hubbard model is investigated
on the square lattice in the thermodynamical limit. The obtained charge
excitation spectra consist of a continuum, a gapless collective mode with
anisotropic zero-sound velocity, and a correlation induced high-frequency mode
at $\omega\approx U$. The correlation function is calculated from Gaussian
fluctuations around the paramagnetic saddle-point within the Kotliar and
Ruckenstein slave-boson representation. Its dependence on the on-site Coulomb
repulsion $U$ and density is studied in detail. An approximate analytical
expression of the high frequency mode, that holds for any lattice
with one atom in the unit cell, is derived. Comparison with numerical 
simulations, perturbation theory and the polarization potential theory is
carried out. We also show that magnetic instabilities tend to vanish for
$T\gtrsim t/6$, and finite temperature phase diagrams are established.
\end{abstract}

\maketitle
\section{Introduction}
\label{sec:int}

Our understanding of excitations in correlated electron systems has been
strongly influenced by the seminal works of Hubbard~\cite{Hub63}, 
Landau~\cite{Lan56}, and Pines~\cite{PinBoh}. In his study~\cite{Hub63} of the 
model Hamiltonian that is now associated with his name, Hubbard put forward one 
very important feature of strongly correlated electrons: the splitting of the 
bands into the so-called upper Hubbard band (UHB) and the lower Hubbard band 
(LHB). Its origin can be traced back to the atomic limit where a gap of the 
order of the interaction strength separates two sets of states, one at 
$\omega\approx 0$ and the other one at $\omega \approx U$. One then expects that by 
ramping up the hopping between the sites, the hybridization of the atomic orbitals 
progressively results in the delocalized states forming the dispersive LHB and 
UHB. However Hubbard's treatment fails to produce the predicted 
Fermi liquid for weak coupling. Indeed, in this regime the self-energy he postulated 
neither reduces to the perturbation theory result, nor does it yield the correct Fermi 
surface for the metallic phase~\cite{Edw68}. In the latter limit Landau's theory of the 
Fermi liquid~\cite{Lan56,PinNoz,Vol84} has proved to be a successful paradigm for 
understanding a large variety of fermion systems at low temperature such as normal 
liquid $\,^3$He, metals or semimetals, and nuclear matter. This phenomenological 
approach is based on the physical intuition that the low-energy properties of 
interacting particles can be modeled from a gas of elementary excitations, 
referred to as quasiparticles, which are formed with a life time that is infinite 
on the Fermi surface but rapidly decays away from it. Alternatively to Landau's 
original formulation this result can be obtained within perturbation theory. 
Using the latter to compute correlation functions within the random phase
approximation (RPA), Pines and Bohm~\cite{PinBoh} showed that the response 
functions are composed of a continuum generated by the incoherent response of
quasiparticles supplemented by peaks signaling collective excitations, that
arise as dynamical fluctuations of the groundstate. The dispersion and 
attenuation of the collective modes are then indicative of the nature of the 
phase. Prominent examples are the Goldstone modes which appear when a continuous 
symmetry is spontaneously broken at a phase transition, such as phonons for 
rotational and translational symmetries, or magnons for the spin-rotational 
symmetry.

Reconciling the Fermi liquid with Hubbard's local physics remains an important 
and largely unsolved problem for correlated electrons. Indeed numerical approaches 
generically face finite-size effects~\cite{Bor91} because required computing 
resources blow up exponentially with increasing system size. 
Yet results can be obtained in limiting cases such as the infinite-coordination 
lattice where the dynamical mean-field theory~\cite{Geo96,Bul01} catches the Hubbard gap 
in the vicinity of half filling. However in the doped system a clear picture is still 
missing, possibly because of the formation of incommensurate phases with large unit cells 
which cannot be captured by the method and its cluster extensions~\cite{Shr90,Fre91,
SeiSi,Sei02,Rac06,Cha10,stripes,Cor14,Lep15}. Furthermore, the above mentioned 
approaches are mainly focused on a self-consistent calculation of one-particle 
correlation functions, which can be directly related to experimental observations 
such as photoemission. However other experimental techniques require the 
knowledge of two-particle quantities such as the charge and spin response 
functions that are probed in neutron scattering experiments. Computing 
two-particle correlations is more challenging because of the need to include 
vertex corrections~\cite{Geo96,Haf14}.   

The purpose of this work is to compute the charge response function of the 
Hubbard model using an extension of the Kotliar and Ruckenstein slave-boson 
representation. One of our main results is that it reduces to the RPA 
susceptibility for weak coupling. 
The obtained charge excitation spectrum generically consists of i) a continuum 
the width of which decreases with increasing interaction strength and density, 
ii) a collective mode with anisotropic zero-sound (ZS) velocity, and iii) a high
frequency mode at $\omega \approx U$ which is the signature of the UHB. Hence 
our scheme reconciles the Fermi liquid physics --- including collective modes 
--- with Hubbard's local physics embedded in the split bands. The calculation 
is carried out in a paramagnetic phase, free of symmetry breaking, in the 
thermodynamical limit. It allows us to resolve the full momentum dependence of the 
spectra. At first glance, neglecting magnetic instabilities puts severe constraints 
on the parameter range where the calculation may be meaningfully performed. However, 
as shown below, the incommensurate magnetic instabilities are strongly suppressed 
with increasing temperature, so that they essentially disappear for $T\approx t/6$.

Since Mott insulating groundstates arise at large $U$ and at half
filling, we perform our investigations in a framework which is able to
capture interaction effects beyond the physics of Slater determinants.
We use an extension of the Kotliar and Ruckenstein slave-boson
representation that reproduces the Gutzwiller approximation on the
saddle-point level~\cite{Kot86}. It entails the interaction driven
Brinkman-Rice metal-to-insulator transition~\cite{Bri70}. A whole range
of valuable results have been obtained with Kotliar and 
Ruckenstein~\cite{Kot86} and related slave-boson 
representations~\cite{Li89,FW}, which motivates the present study. In 
particular they have been used to describe anti-ferromagnetic~\cite{Lil90}, 
spiral~\cite{Fre91,Igo13,Fre92,Doll2}, and striped~\cite{SeiSi,Fle01,Sei02,Rac06} 
phases. Furthermore, the competition between the latter two has been 
addressed as well~\cite{RaEPL}. Besides, it has been obtained that the 
spiral order continuously evolves to the ferromagnetic order in the
large $U$ regime ($U \gtrsim 60t$)~\cite{Doll2} so that it is unlikely to be realized
experimentally. Consistently, in the two-band model, ferromagnetism was
found as a possible groundstate only in the doped Mott insulating 
regime~\cite{Fre02}. Yet, adding a ferromagnetic exchange coupling was 
shown to bring the ferromagnetic instability line into the intermediate
coupling regime~\cite{lhoutellier15}. A similar effect has been obtained
with a sufficiently large next-nearest-neighbor hopping amplitude~\cite{FW98} or
going to the fcc lattice~\cite{Igo15}.

The influence of the lattice geometry on the metal-to-insulator transition 
was discussed, too~\cite{Kot00}. For instance, a very good agreement with
Quantum Monte Carlo simulations on the location of the
metal-to-insulator transition for the honeycomb lattice has been
demonstrated~\cite{Doll3}. Also, strongly inhomogeneous polaronic states 
that have been found in correlated heterostructures have also been 
addressed using this formalism applied to the Hubbard model extended 
with inter-site Coulomb interactions~\cite{Pav06}. Most recently the
approach has been used to address possible capacitance enhancement in
a capacitor consisting of strongly correlated plates separated by a
dielectric~\cite{Ste17}. Furthermore, comparison of groundstate energies 
to existing numerical solutions have been carried out for the square 
lattice, too. For instance, for $U=4t$ it could be shown that the 
slave-boson groundstate energy is larger than its counterpart by less 
than 3\%~\cite{Fre91}. For larger values of $U$, it has been obtained 
that the slave-boson groundstate energy exceeds the exact diagonalization 
data by less than 4\% (7\%) for $U=8t$ ($20t$) and doping larger than 
15\%. The discrepancy increases when the doping is lowered~\cite{Fre92}. 
It should also be emphasized that quantitative agreement to quantum Monte
Carlo charge structure factors was established~\cite{Zim97}.

The paper is organized as follows. In Sec.~\ref{sec:method} we give a 
brief presentation of the spin-rotation-invariant (SRI) Kotliar and 
Ruckenstein slave-boson representation of the Hubbard model and the method 
used to calculate dynamical response functions (more details can be found 
in, e.g., review~\cite{fresard12}). Sec.~\ref{sec:saddle} presents the 
paramagnetic saddle-point solution and discusses its temperature 
dependence. In addition, phase diagrams summarizing the temperature dependence
of magnetic and charge instabilities are established. We evaluate the spin and 
charge susceptibilities from fluctuations captured within the one-loop 
approximation in Sec.~\ref{sec:response} and investigate the dispersion of their 
collective modes in Sec.~\ref{sec:modes}. Our results are discussed in comparison 
with the perturbation Hartree-Fock (HF)+ RPA prediction, as well as with available 
numerical investigations (exact diagonalization and Quantum  Monte Carlo method) 
in Sec.~\ref{sec:comparison}. And we summarize the paper in the conclusion. 

\section{Model and method} \label{sec:method}
The Hubbard Hamiltonian in the SRI Kotliar and Ruckenstein slave-boson 
representation~\cite{Kot86,fresard12} is expressed with auxiliary boson 
operators $e_i$, $p_{i\mu}$, $d_i$ (for atomic states with respectively zero, 
single and double occupancy) and pseudo-fermion operators $f_{i\sigma}$ as
\begin{equation} \label{eq:model}
 H = \sum_{i,j} t_{ij}\sum_{\sigma, \sigma', \sigma''} z_{i\sigma'' \sigma}^{\dagger} 
 f_{i\sigma}^{\dagger} f_{j\sigma'} z_{j\sigma' \sigma''} + U \sum_i d_i^{\dagger} d_i.
\end{equation}
Here hopping occurs between nearest neighbor sites with amplitude 
$t_{ij}=-t$. A key feature of the representation lies in the reduction of 
the on-site Coulomb interaction into a term bilinear in bosonic operators, 
at the expense of a more complicated hopping term. In order to preserve 
spin rotation symmetry~\cite{Li89,FW} the canonical operator $p_{i\mu}$ 
build a $2\times2$ matrix in spin space that is expanded into the 
identity matrix $\underline{\tau}^0$ and the Pauli matrices as 
${\underline p}_i = \frac{1}{2} \sum_{\mu=0}^3 p_{i\mu} {\underline \tau}^{\mu}$. 
In this space the occupancy-change operator $\underline{z}_i$ in the 
hopping term is also a matrix defined as
\begin{equation}
 {\underline z}_i = e_i^{\dagger} {\underline L}_i M_i {\underline R}_i \, 
 {\underline p}_i + {\underline {\tilde{p}}}_i^{\dagger} {\underline R}_i M_i  
 {\underline L}_i \, d_i
\end{equation}
with 
\begin{eqnarray}
& M_i & = \Big[ 1 + e_i^{\dagger} e_i + \sum_{\mu=0}^3 p_{i\mu}^{\dagger} p_{i\mu} 
+  d_i^{\dagger} d_i \Big]^{1/2}, \nonumber \\
&{\underline L}_i & = \Big[ (1 -d_i^{\dagger} d_i) {\underline \tau}^0 
- 2 {\underline p}_i^{\dagger} {\underline p}_i \Big]^{-1/2}, \nonumber \\  
& {\underline R}_i & = \Big[ (1 - e_i^{\dagger} e_i) {\underline \tau}^0 
- 2 {\tilde{\underline p}}_i^{\dagger} {\tilde{\underline p}}_i \Big]^{-1/2}          
\end{eqnarray}
where  $\tilde{\underline p}_i = \frac{1}{2} ( p_{i0} {\underline \tau}^0 
- {\bf p}_i \cdot \boldsymbol{\underline \tau} )$.

In the augmented Fock space generated by the auxiliary boson operators, 
the subspace of physical states is the intersection of the kernels of 
operators 
\begin{eqnarray}\label{eq:const}
 & {\cali A}_i &= e_i^{\dagger} e_i + \sum_{\mu=0}^3 p_{i\mu}^{\dagger} p_{i\mu} 
 +  d_i^{\dagger} d_i - 1,  \nonumber\\
 & {\cali B}_{i0} & = \sum_{\mu=0}^3 p_{i\mu}^{\dagger} p_{i\mu} + 2 d_i^{\dagger} d_i 
 - \sum_{\sigma} f_{i \sigma}^{\dagger} f_{i \sigma}, \\
 & \pmb{\cali B}_i & =  p_{i0}^{\dagger} {\bf p}_i + {\bf p}_i^{\,\dagger} p_{i0} 
 - \ii {\bf p}_i^{\,\dagger} \times {\bf p}_i  - \sum_{\sigma, \sigma'} 
 \boldsymbol{ \tau}_{\sigma \sigma'} f_{i\sigma'}^{\dagger} f_{i\sigma}, \nonumber
\end{eqnarray}
{\it i.e.} in this subspace ${\cali A}_i = 0$ that is the constraint of one 
atomic state per site, and ${\cali B}_{i\mu} = 0$ which equates the 
number of fermions to the number of $p$ and $d$ bosons.

The partition function is calculated as a functional integral~\cite{li91,Zim97} 
with the effective Lagrangian 
${\cal L} = {\cal L}^{\rm B} + {\cal L}^{\rm F}$ where the purely 
bosonic part is
\begin{eqnarray}
 {\cal L}^{\rm B} = \sum_i  \bigg[  e_i^{\dagger} \partial_{\tau} e_i
 & + & \sum_{\mu=0}^3 p_{i\mu}^{\dagger} \partial_{\tau} p_{i\mu}  
 + d_i^{\dagger} (\partial_{\tau} + U) d_i  \nonumber \\
 & + & \alpha_i {\cali A}_i + \sum_{\mu=0}^3 \beta_{i\mu}  {\cali B}_{i\mu}^{\rm B} \bigg]
\end{eqnarray}
with ${\cali B}_{i\mu}^{\rm B}$ being the bosonic part of the operator 
${\cali B}_{i\mu}$, and the mixed-fermion-boson part can be written as
\begin{eqnarray}
{\cal L}^{\rm F} = & - {\rm tr} \Big\{ \ln \Big[ (\partial_{\tau} - \mu +  \beta_{i0}) 
\delta_{\sigma\sigma'} \delta_{ij} + \boldsymbol{\beta}_i \cdot 
\boldsymbol{\tau}_{\sigma\sigma'} \delta_{ij} \nonumber \\
 & + t_{ij} \sum_{\sigma_1} z_{j\sigma\sigma_1}^{\dagger} z_{i\sigma_1\sigma'} \Big] \Big\}
\end{eqnarray}
after the fermion fields have been integrated (here $\mu$ is the chemical 
potential). The constraints that define the physical states are enforced 
with Lagrange multipliers $\alpha_i$ and $\beta_{i\mu}$. The internal gauge 
symmetry group of the representation allows to simplify the problem. The 
phases of $e$ and $p_{\mu}$ can be gauged away by promoting the Lagrange 
multipliers to time-dependent fields~\cite{FW}, leaving us with radial 
slave-boson fields~\cite{Fre01}. Their values obtained at the saddle-point level 
may be viewed as an approximation to their exact expectation values that are 
generically non-vanishing~\cite{Kop07}. The slave-boson field 
corresponding to double occupancy $d_i = d'_i + {\rm i} d''_i$ however 
has to remain complex as emphasized by several authors~\cite{Jol91,Kot92,FW}. 
Since $e_i$ and $p_{i\mu}$ are now real, their kinetic terms drop out of 
${\cal L}^{\rm B}$ due to the periodic boundary conditions on boson fields.

Within the approximation of Gaussian fluctuations, the action is 
expanded to second order in field fluctuations 
\begin{eqnarray}
&\psi(k) = \big(\delta e(k),\delta d'(k),\delta d''(k),\delta p_{0}(k),
\delta \beta_{0}(k), \delta \alpha(k), \nonumber \\ 
& \delta p_{1}(k),\delta \beta_{1}(k),\delta p_{2}(k),\delta \beta_{2}(k),
\delta p_{3}(k), \delta \beta_{3}(k) \big)
\end{eqnarray}
around the paramagnetic saddle-point solution 
\begin{equation}
\psi_{\rm MF} = (e, d, 0, p_0,\beta_0,\alpha, 0,0,0,0,0,0)
\end{equation}
as
\begin{equation}
 \int d\tau {\cal L}(\tau) = {\cal S}_{\rm MF} + \sum_{k,\mu,\nu} \psi_{\mu}(-k) 
 S_{\mu\nu}(k) \psi_{\nu}(k)
\end{equation}
(the matrix $S$ is given in Appendix~\ref{app_S}). We have introduced the 
notation $k=({\bf k},\nu_n)$, where $\nu_n=2\pi nT$, and 
$\sum_k = T \sum_{\nu_n} L^{-1} \sum_{{\bf k}}$ with $L$ the number of 
lattice sites. The correlation functions of boson fields are then 
Gaussian integrals which can be obtained from the inverse of the fluctuation 
matrix $S$ as 
$\langle \psi_{\mu}(-k) \psi_{\nu}(k) \rangle = \frac{1}{2} S^{-1}_{\mu\nu}(k)$. 
For instance the slave-boson representation of the spin fluctuation 
$\delta {\cali S}_z = \delta(p_0^{\dagger} p_3 + p_3^{\dagger} p_0)$ 
yields the spin susceptibility
\begin{equation}
 \chi_s(k)  =  \langle \dd {\cali S}_z(-k) \dd {\cali S}_z(k) \rangle = 2 p_0^2 
 S_{11,11}^{-1}(k) .
\end{equation}
Similarly, using the density fluctuation 
$\delta {\cali N} = \delta(d^{\dagger} d - e^{\dagger} e) $, 
the charge susceptibility is
\begin{eqnarray}
 \chi_c(k) & = & \langle \dd {\cali N}(-k) \dd {\cali N}(k) \rangle \nonumber \\
 &  = & 2 e^2 S_{1,1}^{-1}(k) - 4 e d S_{1,2}^{-1}(k) + 2 d^2  S_{2,2}^{-1}(k) .
 \label{eq:chi_c-def}
\end{eqnarray}
Dynamical response functions are eventually evaluated within analytical 
continuation ${\rm i} \nu_n \rightarrow \omega + {\rm i} 0^+$.

The saddle-point approximation is exact in the large degeneracy limit, 
and the Gaussian fluctuations provide the $1/N$ corrections~\cite{FW}. 
Moreover it obeys a variational principle in the limit of large spatial 
dimensions where the Gutzwiller approximation (GA) becomes exact for the 
Gutzwiller wave function~\cite{Met89}.

\section{Paramagnetic saddle-point solution} \label{sec:saddle}

\subsection{Characterization of the paramagnetic phase}

\begin{figure}[b]
   \includegraphics[clip=true, width=0.44\textwidth]{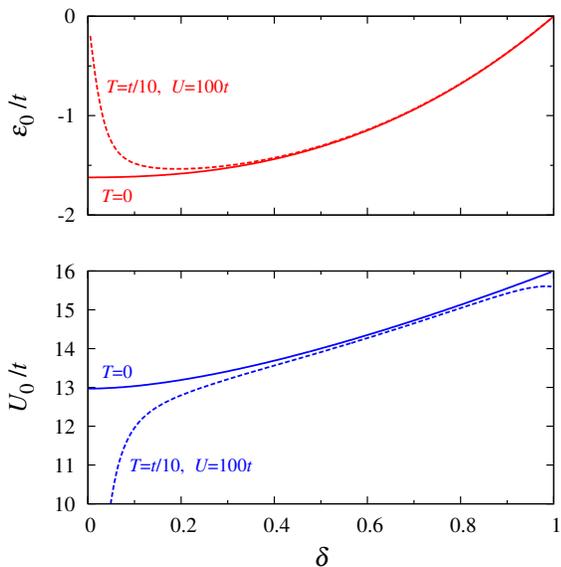}
	\caption{(Color online) Doping dependence of kinetic energy $\varepsilon_0$ 
	and coupling scale $U_0$ in the metallic phase at $T=0$ and for $U=100t$ 
	at $T=t/10$.}
	\label{fig:eps0_U0}
\end{figure}

At the paramagnetic saddle-point, the field $\underline{z}_i$ reduces to 
$z_0 \underline{\tau}^0$ with
 \begin{equation}
 z_0=p_0(e+d)\sqrt{\frac{2}{1-\delta^2}}
 \end{equation}
where $\delta = 1 - \langle {\cali N} \rangle$ is the hole doping from 
half-filling. The factor $z_0^2$ plays the role of a quasiparticle 
residue, and it also renormalizes the quasiparticle dispersion as 
\begin{equation}
 E_{{\bf k}} = z_0^2 t_{{\bf k}} - (\mu - \beta_0)
\end{equation}
with the bare dispersion $t_{{\bf k}} = -2t(\cos k_x +\cos k_y)$ for the 
square lattice. 

\begin{figure}[b]
	 \includegraphics[clip=true, width=0.45\textwidth]{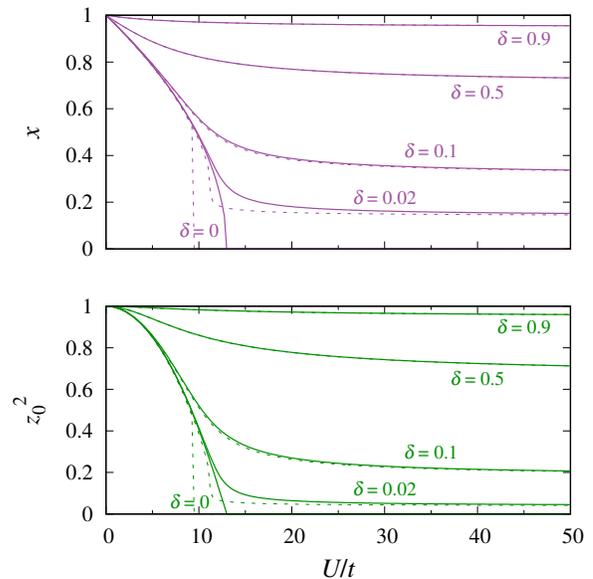}
	\caption{(Color online) Saddle-point variable $x$ and renormalization 
	factor $z_0^2$ for different values of coupling $U$ and doping $\delta$ 
	at $T=0$ (solid line) and at $T=t/10$ (dashed line).}
	\label{fig:x-z2}
\end{figure}

The boson saddle-point values can be expressed with the 
doping and the variable $x =e +d$ as  
\begin{eqnarray}
& e = \frac{x^2 + \delta}{2x}, \quad d = \frac{x^ 2 - \delta}{2 x} ,
\quad p_0^2 = 1 - \frac{x^4 + \delta^ 2}{2 x^2} , \\
& \alpha  =  \frac{p_0^2 x^2 U_0}{2} \left(\frac{1}{x^2 + \delta} + \frac{1}{1-\delta} 
\right), \; \beta_0  =  \alpha - \frac{x^2 U_0}{4} \left( 1 + \frac{2 p_0^2}{1-\delta^2} 
\right) \nonumber
\end{eqnarray}
(the expressions result from the constraints on physical states 
$e^2 + p_0^2 + d^2 = 1$ and $p_0^2 + 2 d^2 = 1- \delta$, and saddle-point 
conditions). Here the coupling scale
\begin{equation}
 U_0 = - 8 \varepsilon_0/(1-\delta^2)
\end{equation}
has been introduced in terms of the semi-renormalized kinetic energy 
 \begin{equation}
 \varepsilon_0 =  \frac{2}{L} \sum_{{\bf k}}  t_{{\bf k}} n_F(E_{{\bf k}})
 \end{equation}
 and the Fermi function $n_F(\epsilon) = 1/(\exp(\epsilon/T) + 1)$.

As discussed in Ref.~\cite{Vol87,FW}, the paramagnetic solution for fixed 
values of doping $\delta$ and coupling $U$ is found by determining 
the chemical potential via the filling condition
\begin{equation}
  \frac{2}{L} \sum_{{\bf k}} n_F(E_{{\bf k}}) = 1 - \delta
\end{equation}
 and the solution of the saddle-point equation
\begin{equation}
\frac{(1-x^2) x^4}{x^4 - \delta^2}  =  \frac{U}{U_0}.
\end{equation}
The procedure is carried out self-consistently with the evaluation of 
$z_0$ since the latter renormalizes the dispersion. It is however 
simplified at $T=0$ because then, for a fixed filling, $\varepsilon_0$ 
and $U_0$ have the same values for all finite $z_0$. This implies they 
do not vary with the coupling, except at $\delta=0$ where they vanish 
discontinuously above a critical coupling $U_c$. Fig.~\ref{fig:eps0_U0} 
displays their variations with the doping. Increasing the temperature
from zero reduces their amplitudes and it smooths out the discontinuity 
at half filling while enlarging the collapse around it, as shown by the 
curves plotted at temperature $T=t/10$.

\begin{figure}[h]
	 \includegraphics[clip=true, width=0.45\textwidth]{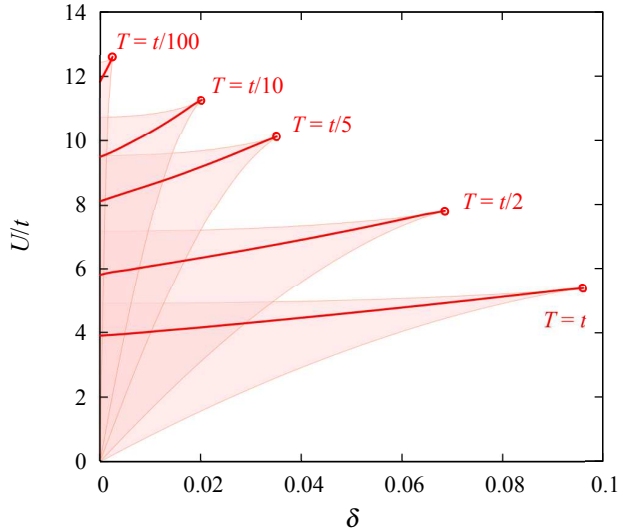}
	\caption{(Color online) Region of the phase diagram with a degeneracy 
	of the saddle-point solution at different temperatures. The solid line 
	with its critical endpoint indicates the metal-to-bad-metal 
	first-order transition taking place when increasing the coupling.}
	\label{fig:MFdegeneracy}
\end{figure}

For most values of coupling and doping, the saddle-point equation 
possesses one finite solution $x>0$ corresponding to a metallic state. 
As shown in Fig.~\ref{fig:x-z2} saddle-point values converge in the 
infinite coupling limit where $x=\sqrt{|\delta|}$ and 
$z_0^2 = 2|\delta|/(1+|\delta|)$. A remarkable phenomenon occurs at 
half filling where $x$ vanishes above the critical coupling that is 
$U_c = - 8 \varepsilon_0=2(8/\pi)^2 t \approx 12.97 t$ at $T=0$. This 
solution corresponds to an insulating state since $z_0^2=0$ results 
in a diverging quasiparticle mass and a vanishing quasiparticle 
residue: This is the Brinkman-Rice mechanism~\cite{Bri70} for the 
Mott metal-to-insulator transition. Note that at finite temperature, 
for small doping and $U<U_c$, the equation can have up to three 
positive solutions~\cite{Doll3}, among which the groundstate is 
determined by minimizing the free energy
\begin{eqnarray}
 F & = & \Omega + \mu  \langle {\cali N} \rangle \\
 & = & - \frac{2 T}{L} \sum_{{\bf k}} \ln \left[1 + \mathrm{e}^{-E_{{\bf k}}/T}\right] 
 + U d^2 + (\mu - \beta_0) (1-\delta). \nonumber
\end{eqnarray}
The degeneracy of solutions gives rise to a first-order transition 
when increasing the coupling from a metallic state into either an 
insulating state at half filling~\cite{FK97}, or a bad metallic one 
characterized by a small quasiparticle residue $z_0^2$ for finite 
doping (see Fig.~\ref{fig:x-z2}). Fig.~\ref{fig:MFdegeneracy} shows 
the transition line in the $(\delta,U)$-phase diagram at different 
temperatures. It is terminated by a critical endpoint at finite doping, 
and as well as the parameter region with multiple solutions, it shrinks 
with lowering temperature and vanishes at $T=0$~\cite{Doll3,camjayi07}. 
Contrary to an ordinary band insulator where thermal excitations of 
quasiparticle enhance the conductivity, increasing the temperature in 
the strongly correlated Hubbard model can induce a transition from a 
low-temperature metal to a high-temperature insulator as thermal 
fluctuations destroy the poor coherence of the small-$z_0$ metallic 
state in a fashion similar to the transition observed in 
V$_2$O$_3$~\cite{McWhan73}. 

\subsection{Instabilities}

\begin{figure}[h!]
	 \includegraphics[clip=true, width=0.42\textwidth]{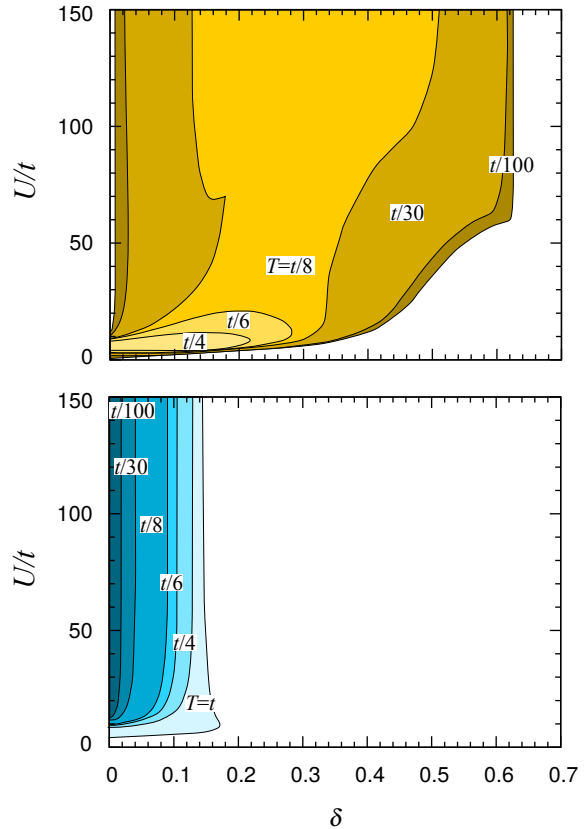}
	\caption{(Color online) Instability of the paramagnetic phase toward 
	incommensurate magnetic ordering (top) or phase separation (bottom), 
	at different temperatures. 
	The shaded area corresponds to values of doping and coupling for which 
	the static spin/charge susceptibility diverges.}
	\label{fig:phasediagram}
\end{figure}

Let us now look for the parameter range in which the above solution is stable. 
In Ref.~\cite{doll93} it was found that the zero-temperature slave-boson 
paramagnetic phase is stable at low density, even at large coupling, while  
incommensurate magnetic instabilities develop at large densities. One then may 
ask what is the picture at finite temperature, especially since early estimates 
at half-filling yield a temperature at which magnetic instabilities are destroyed 
to be of order $t^2/U$~\cite{Doll3}. Furthermore, while there is a regain 
of interest in charge instabilities at $T=0$~\cite{lhoutellier15,Ste17}, little 
attention has been paid to them at finite temperature. 
Hence the robustness of the saddle-point solution against spin and charge 
fluctuations is investigated by looking for a divergence of the 
respective static response functions (see Eq.~(\ref{eq:chi_s}) and 
Eq.~(\ref{eq:chi_c})). 
The instabilities of the paramagnetic phase at 
different temperatures are mapped in Fig.~\ref{fig:phasediagram}. The 
static spin susceptibility $\chi_s({\bf k},\omega=0)$, given by 
Eq.~(\ref{eq:chi_s}), has no pole at high temperature but a magnetic 
instability appears below $T \approx t$, around half filling 
and for a finite but not large coupling. Its domain in the 
$(\delta,U)$-phase diagram then grows with lowering temperature, with 
a significant variation between $T=t/6$ and $t/8$. Earlier 
studies~\cite{Fre92,doll93} have found that the instability boundary 
in the phase diagram at $T=0$ signals a magnetic ordering into a spiral 
groundstate. The doping range of the magnetic phase increases with the 
coupling up to the maximum doping $\delta \approx 0.63$ reached at 
$U\gtrsim 60 t$. Contrary to the magnetic behavior, the domain of the 
charge instability shrinks with lowering temperature. It is limited to 
small doping and occurs at all coupling above a moderate threshold 
value which increases with lowering temperature. The charge instability 
is related to a tendency towards a phase separation~\cite{Fre92} or 
towards the more complicated stripe phases~\cite{SeiSi,Sei02,Rac06}. 

\section{Expressions of the dynamical response functions} \label{sec:response}

The evaluation of correlation functions is simplified in the paramagnetic state 
because the Gaussian fluctuations decouple into spin and charge channels. This 
results into a matrix $S$ that is block diagonal with a charge $6\times 6$ submatrix 
and three identical $2 \times 2$ blocks for the components of the spin. As discussed 
in, e.g., Refs.~\cite{Lav90,Zim97,li91,li94}, the blocks can be independently inverted to 
yield the spin (see Appendix~\ref{app_chi_s}) and the charge dynamical response 
function
\begin{widetext}
\begin{equation}
 \chi_c(k) = \frac{e^2 S_{55}(k) \big[  \tilde{S}_{33} \big(2 p_0^2 \Gamma_1(k) 
 - 8 d p_0 \Gamma_2(k) + 8 d^2 \Gamma_3(k) \big) + 2 e^2 p_0^2 S_{55}(k) 
 (\omega + {\rm i} 0^+)^2 \big]} {\tilde{S}_{33} \big[\Gamma_2^2(k) - \Gamma_1(k) \Gamma_3(k) \big] 
 -  \frac{e^2}{(e+d)^2} S_{55}(k) \big[ p_0^2  \Gamma_1(k) + 2 (e-d)p_0 \Gamma_2(k) 
 + (e-d)^2 \Gamma_3(k) \big] (\omega + {\rm i} 0^+)^2}
 \label{eq:chi_c}
\end{equation}
\end{widetext}
with
\begin{eqnarray}
 \tilde{S}_{33} &=& -\frac{ 2 e p_0^2 }{d (1 - \delta^2)}  \varepsilon_0, 
 \nonumber \\
 \Gamma_1(k) & = & -S_{55}(k) [ e^2 S_{22}(k) - 2 e d S_{12}(k) + d^2 S_{11}(k) ] 
 \nonumber \\
& & + [e S_{25}(k) - d S_{15}(k)]^2, \nonumber \\
 \Gamma_2(k) & = & -S_{55}(k) [ e^2 S_{24}(k) - p_0 e S_{12}(k) - e d S_{14}(k) 
 \nonumber \\
& & + d p_0 S_{11}(k) ] + [e S_{25}(k) - d S_{15}(k)] \nonumber \\
& & [e S_{45}(k) - p_0 S_{15}(k)], \nonumber \\
 \Gamma_3(k) & = & -S_{55}(k) [ e^2 S_{44}(k) - 2 e p_0 S_{14}(k) + p_0^2 S_{11}(k) ] 
 \nonumber \\
& & + [e S_{45}(k) - p_0 S_{15}(k)]^2.
\end{eqnarray}
The susceptibilities are particle-hole symmetric as expected for the Hubbard 
model on the square lattice.

The expression of $\chi_c(k)$ given in Ref.~\cite{Zim97} is valid only at zero 
frequency because the matrix elements omitted in the previous work vanish in 
the static limit. 
We have checked that the numerical discrepancies between the charge 
structure factors evaluated in Ref.~\cite{Zim97} and using Eq.~(\ref{eq:chi_c}) 
are minor. They do not alter the previous conclusion that slave-boson results 
are in very good agreement with Quantum Monte Carlo calculations~\cite{Zim97,DziUnpub}. 
However the missing matrix elements are crucial in the investigation of charge 
collective modes. Without them, the poles of the dynamical response function (or 
their residues) would not vanish in the free-particle limit. Furthermore their 
dispersions would depend on the sign of the doping, which is in conflict with the 
particle-hole symmetry expected for the Hubbard model on a bipartite lattice.

\begin{figure}[b]
	 \includegraphics[clip=true, width=0.47\textwidth]{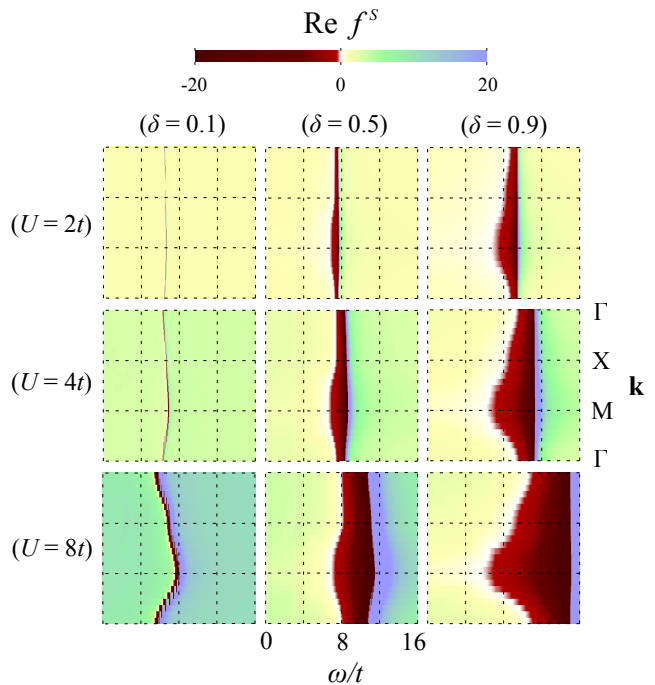}
	\caption{(Color online) Real part of the function $f^s(k)$ at temperature 
	$T=t/100$ for different values of coupling and doping,  plotted for momenta 
	along the path linking ${\rm \Gamma}=(0,0)$, ${\rm X}=(\pi,0)$ 
	and ${\rm M}=(\pi,\pi)$.  }
	\label{fig:fs}
\end{figure}

In the weak-coupling limit slave-boson expressions yield the textbook 
results derived from perturbation methods. This is obtained by 
writing the charge susceptibility~(\ref{eq:chi_c}) as
\begin{equation}
 \chi_c(k) = \frac{\chi_0(k)}{1 + f^s(k) \chi_0(k)}
  \label{eq:chi_c_weak}
\end{equation}
then by expanding the function $f^s(k)$ in powers of the coupling. Here 
the Lindhard function $\chi_0(k)$, given by Eq.~(\ref{eq:chi_m}), 
solely differs from the charge response function of a Fermi gas 
through the quasiparticle mass renormalization $z_0^2$. The function $f^s(k)$ 
is related to the Landau parameter of Fermi-liquid theory~\cite{lhoutellier15} 
by $F_0^s = \chi_0(0) f^s(0) = N_{\rm F} f^s(0)$ where $N_{\rm F}$ is the 
density of states at the Fermi level. Its expansion to first order 
$f^s(k) = U/2$ is in perfect correspondence with the expected RPA result. This 
generalizes Li {\it et al.}'s results~\cite{li91,li94} to arbitrary momentum 
and frequency. Including the next order in $U$ yields
\begin{eqnarray}
 f^s(k) = \frac{U}{2} \bigg[ 1 & + & \frac{U}{2 U_0} \bigg( 4 - (1-\delta^2)
 \big[1+\delta \gamma(k)\big]  \nonumber \\
 & + & \frac{U_0^2}{16} \frac{\big[(1-\delta^2) \gamma(k) - 8 \delta\big]^2}
 {(\omega + {\rm i} 0^+)^2 - (U_0/2)^2} \bigg) \bigg]  
\label{eq:f0s}
\end{eqnarray}
with the ratio $\gamma(k)=\chi_1(k)/\varepsilon_0\chi_0(k)$ and
\begin{equation}
  \chi_m(k) = \frac{2}{L} \sum_{\bf q} (t_{\bf q} + t_{{\bf q} + {\bf k}} )^m 
 \frac{n_F(E_{{\bf q} + {\bf k}}) - n_F(E_{{\bf q} })}
 {(\omega + {\rm i} 0^+) - (E_{{\bf q} + {\bf k}} - E_{\bf q})}.
 \label{eq:chi_m}
\end{equation}
The ratio $\gamma(k)$ in the second order expansion has a complex value. 
Hence the function $f^s(k)$ actually possesses an imaginary part. 
Its real part becomes negative just below a critical energy at which it 
diverges (see Fig.~\ref{fig:fs}). The domain with ${\rm Re} f^s <0$ is largest 
around M. Its size increases with the doping and the coupling. As can be inferred 
from the structure of $f^s(k)$, we show in the next section that the charge 
susceptibility~(\ref{eq:chi_c}) has a rich spectrum that cannot be captured 
within the conventional HF~+~RPA framework.

A theory going beyond the Landau Fermi liquid model and the RPA approximation
has been developed by Pines and coworkers~\cite{Pin81} for the excitations 
and transport properties of quantum liquids. The so-called polarization 
potential (PP) theory is a semi-phenomenological approach that describes the 
collective action of the particles by an averaged self-consistent field which 
can be polarized by particle-hole excitations via an effective screened 
potential.
Using parameters obtained from static measurements and sum rule considerations, 
it attempts to describe both liquid $^4$He and $^3$He within a unified formalism. 
In particular, the theory can reproduce the experimental dispersion of the ZS 
collective mode, beyond the Landau Fermi liquid regime. They obtained a density 
response of the form
\begin{equation}
 \chi_{{\rm pp}}(k) = \frac{\chi^{\rm sc}(k)}{1 + \big[f^s_{\rm pp}({\bf k}) 
 + (\omega^2/{\bf k}^2) g^s_{\rm pp}({\bf k}) \big] \chi^{\rm sc}(k)}
 \label{eq:chi_c_pp}
\end{equation}
within the linear response theory. Two contributions enter the PP. 
The first function $f^s_{\rm pp}({\bf k})$ is the Fourier transform of the  
potential of an effective static particle interaction. The second term corresponds 
to the effect of the so-called backflow, that is the additional screening caused 
by longitudinal current fluctuations accompanying the density fluctuations. In the 
long wavelength limit these quantities are related to the Landau parameters by 
$f^s_{\rm pp}(0) = F_0^s/N_{\rm F} $ and 
$g^s_{\rm pp}(0) = m F_1^s/3\langle {\cali N} \rangle$ where $m$ is the particle 
mass. A reasonable description of the neutron-scattering data on $^4$He and $^3$He 
can be obtained by assuming the PP to be essentially the same for both liquids. The 
influence of the statistics is mainly present in the screened density response 
function $\chi^{\rm sc}(k)$. Using a sum rule argument, the latter is defined as 
the weighted sum of the expression for a free Bose or Fermi gas of particles with 
an effective mass $m^{*}$, and a structureless multiparticle contribution that is 
fitted to the experimental data.

Comparing expression~(\ref{eq:chi_c_weak}) of the density response function with 
equation~(\ref{eq:chi_c_pp}), one can note two distinguishing features. First the PP 
used in $\chi_{\rm pp}(k)$ appears to be an expansion of the function $f^s(k)$ to 
second order in the frequency. With such a frequency dependence, the PP is not singular 
and $\chi_{\rm pp}(k)$ possesses one single pole corresponding to the ZS mode. It
cannot then produce a second collective excitation, contrary to our result. As shown 
by Eq.~(\ref{eq:f0s}), $f^s(k)$ can diverge and it can then give rise to another 
collective mode. As shown below, for strong coupling, the latter disperses around
$\omega\approx U$ and we therefore call it the UHB mode. However the PP theory includes 
a phenomenological multiparticle contribution in the screened density response function 
$\chi_{\rm sc}(k)$, which is absent from the approximation level used in the present 
work. Multiparticle processes may have a significant influence on the collective 
modes as, for instance, within the PP theory they soften the ZS mode at large 
wavevectors. Including them in our approach could, in principle, be achieved with an 
expansion of the action going beyond the Gaussian fluctuation approximation.

\section{Charge collective modes} \label{sec:modes}

\begin{figure}[b]
	 \includegraphics[clip=true, width=0.46\textwidth]{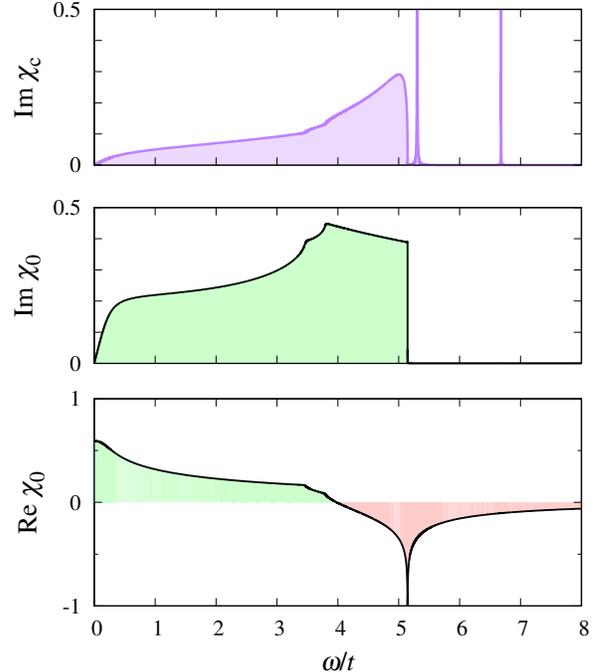}
	\caption{(Color online) Frequency dependence of the charge susceptibility 
	$\chi_c(k)$ and the Lindhard function $\chi_0(k)$ for ${\bf k}=(\frac{\pi}{2},
	\frac{\pi}{2})$ at doping $\delta=0.1$, coupling $U=4t$ and temperature $T=t/100$. 
	With $z_0^2\approx 0.91$ the particle-hole continuum ends at $\omega_{\rm cont}
	({\bf k}) \approx 5.14 t$. }
	\label{fig:chi-U_4}
\end{figure}

\begin{figure}
\begin{tabular}{@{}l@{}}
 \includegraphics[trim=1.55cm 1.75cm 2cm 1.5cm, clip=true, width=0.37\textwidth]
 {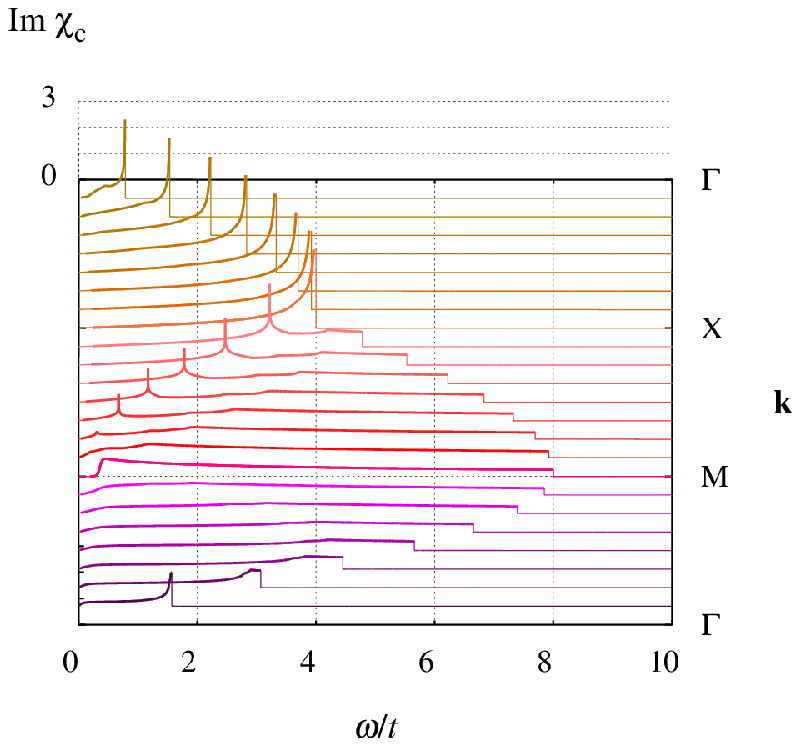} \\
 \includegraphics[trim=1.55cm 1.75cm 2cm 1.5cm, clip=true, width=0.37\textwidth]
 {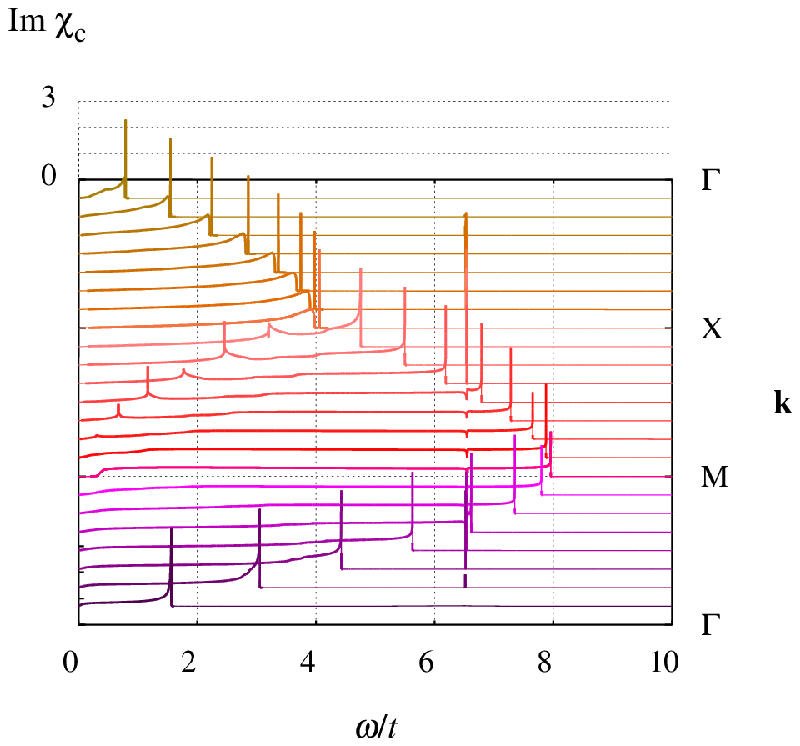} \\
 \includegraphics[trim=1.55cm 1.75cm 2cm 1.5cm, clip=true, width=0.37\textwidth]
 {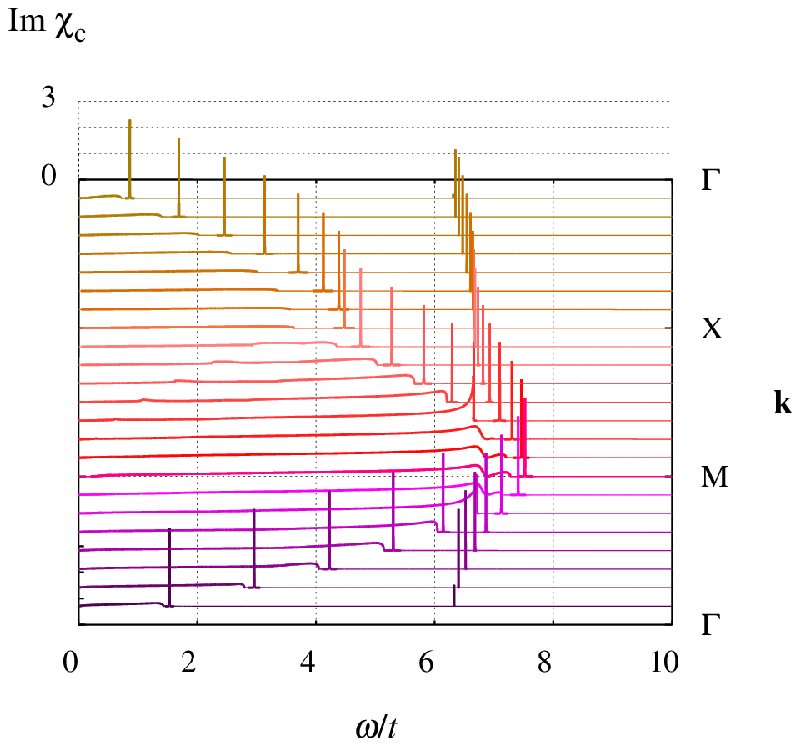} \\
 \includegraphics[trim=2.2cm 0.7cm 2.4cm 1.5cm, clip=true, width=0.475\textwidth]
 {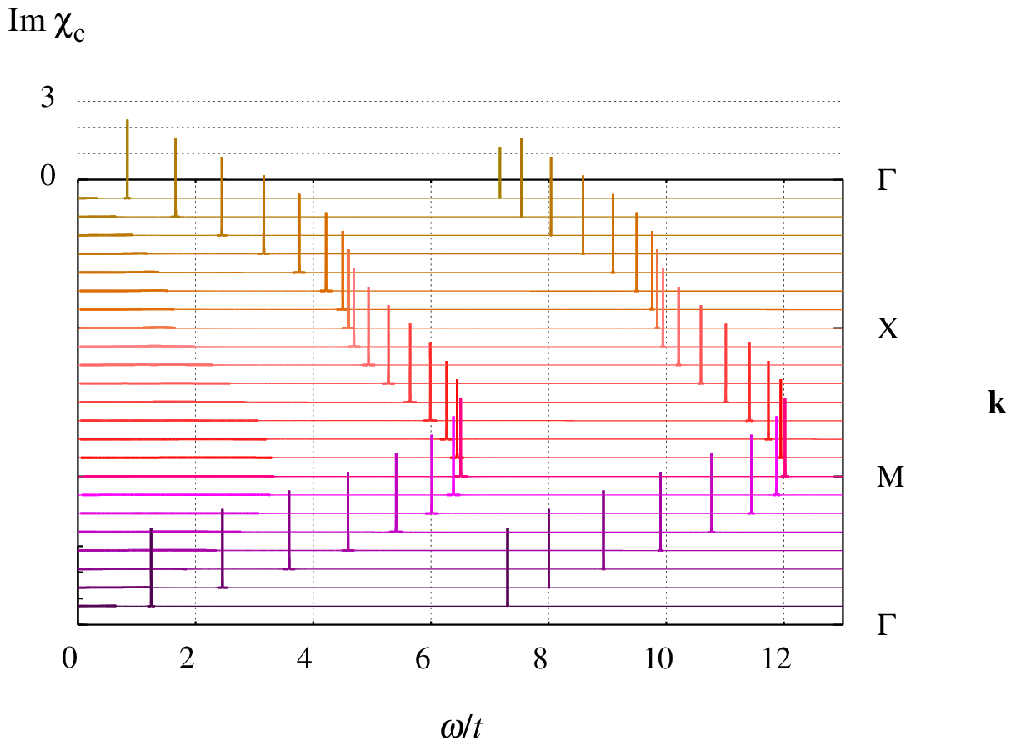}
\end{tabular}
	\caption{(Color online) Imaginary part of the charge susceptibility for   
	$U/t=0$, 1, 4, and 12 from top to bottom, plotted for momenta along the 
	path linking ${\rm \Gamma}=(0,0)$, ${\rm X}=(\pi,0)$ and ${\rm M}=(\pi,\pi)$. 
	Parameters: $T=t/100$ and $\delta=0.1$.}
	\label{fig:chi_c-d_0.1}
\end{figure}

\begin{figure}[b]
\begin{tabular}{@{}l@{}}
 \includegraphics[trim=1.8cm 1.8cm 2.3cm 0.7cm, clip=true, width=0.42\textwidth]
 {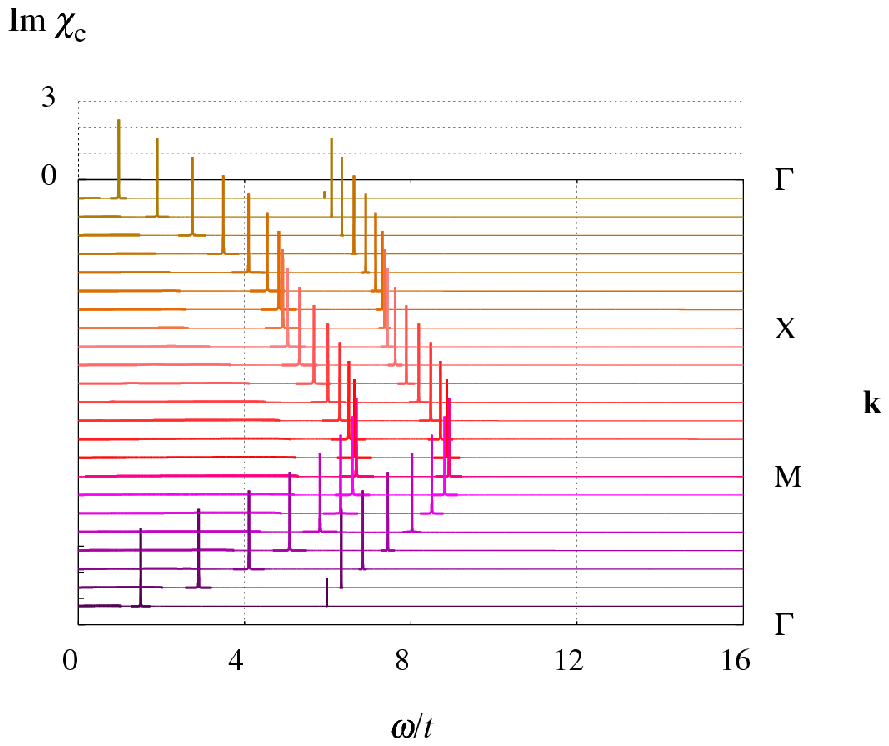} \\
 \includegraphics[trim=1.8cm 1.8cm 2.3cm 1.5cm, clip=true, width=0.42\textwidth]
 {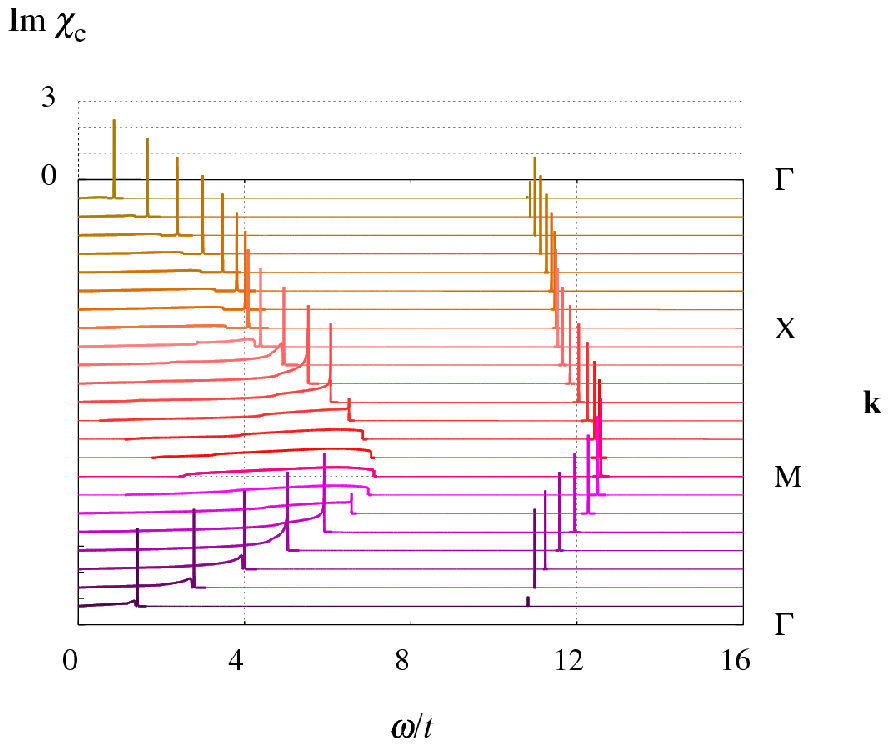} \\
 \includegraphics[trim=1.8cm 0.7cm 2.3cm 1.5cm, clip=true, width=0.42\textwidth]
 {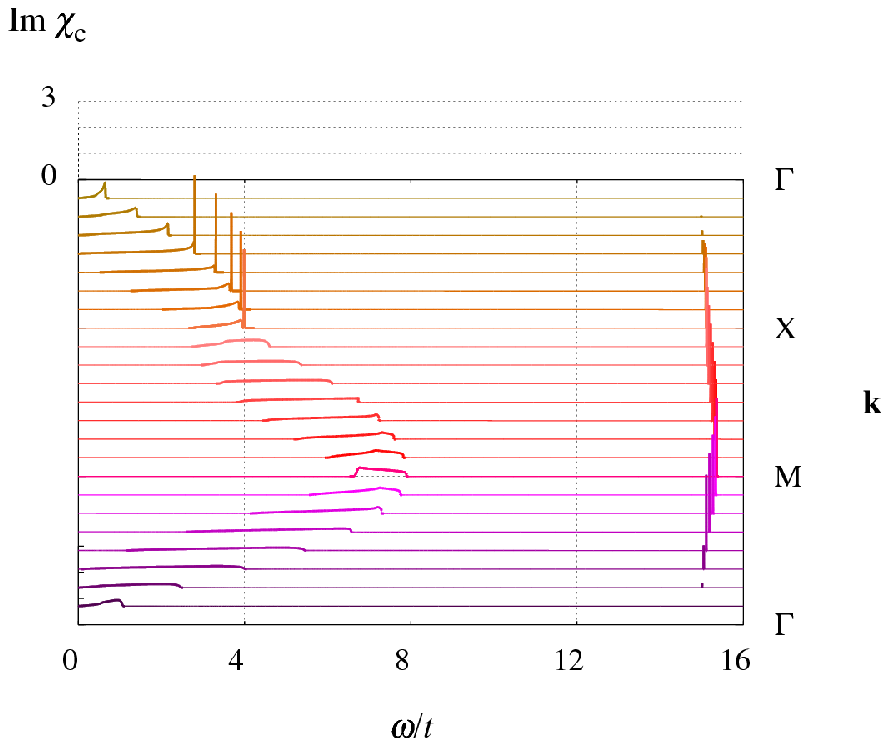}
\end{tabular}
	\caption{(Color online) Imaginary part of the charge susceptibility for 
	doping values $\delta=0.1$, 0.5, and 0.9 from top to bottom. Parameters:  
	$T=t/100$ and $U/t=8$.}
	\label{fig:chi_c-U_8}
\end{figure}

The charge susceptibility possesses two collective modes that appear at finite coupling. 
These excitations form narrow peaks at well defined energies in the spectrum of the 
inelastic response ${\rm Im} \chi_c(k)$. As shown in Fig.~\ref{fig:chi-U_4} the spectrum 
is composed of a broad continuum that results from incoherent single-particle excitations. 
Beyond its upper boundary $\omega_{\rm cont}({\bf k})$ lie the peaks of the two modes. 
The typical evolution  of the charge response function with the coupling is plotted in 
Fig.~\ref{fig:chi_c-d_0.1} and the effect of doping is shown in Fig.~\ref{fig:chi_c-U_8}. 
The continuum contribution to $\chi_c(k)$ is roughly reduced by a factor 
$\sim (1+U N_{\rm F}/2)$  while its energy width shrinks as $z_0^2$. The mode at lower 
energy $\omega_{\rm ZS}({\bf k})$ is the zero-sound mode. It has a linear dispersion at 
long wavelength, that is around the k-point $\Gamma$. It appears as a resonance at the 
upper edge $\omega_{\rm cont}({\bf k})$ and it changes into a well defined peak that 
departs from the continuum when increasing the coupling. The second mode is the 
upper-Hubbard-band mode which occurs at higher energy $\omega_{\rm UHB}({\bf k})$. It 
appears at small coupling with no dispersion at $\omega=U_0/2$ and it then develops with 
a gap at ${\bf k}= \Gamma$ that grows as $U$ in the strong-coupling limit. 

The dispersions of the collective modes are presented below in more details. Since 
our results are best understood at $T=0$ we postpone the discussion of temperature 
effects to the end of the section.

\subsection{Zero-sound mode}

\begin{figure}[b]
  \includegraphics[clip=true, width=0.49\textwidth]
  {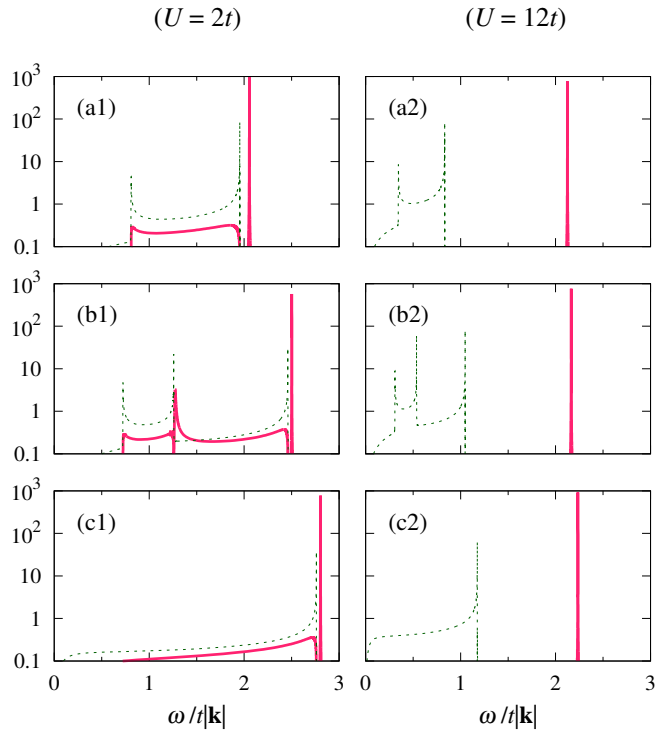}
  \caption{(Color online) Imaginary part of $\chi_c(k)$ (solid line) and of 
  $\chi_0(k)$ (dotted line) for small momentum ${\bf k}= 0.001 (\cos \theta, 
  \sin \theta)$ with angle $\theta =0$ (a), $\pi/10$ (b), and $\pi/4$ (c),
  at moderate coupling $U=2t$ and strong coupling $U=12t$.  
  Parameters: $\delta=0.1$ and $T=0$.}
  \label{fig:zsspectrum}
\end{figure}

The conditions under which the collective modes develop can be discussed with 
the weak-coupling expressions~(\ref{eq:chi_c_weak}) and (\ref{eq:f0s}) for the 
susceptibility. To first order in the coupling, $f^s(k) \approx U/2$ so the 
denominator of $\chi_c(k)$ can vanish only if $\chi_0(k)$ is real and negative. 
As shown in Fig.~\ref{fig:chi-U_4}, these conditions are met beyond the upper 
edge $\omega_{\rm cont}({\bf k})$ of the response continuum, which corresponds 
to the largest energy of the particle-hole excitations with momentum ${\bf k}$. 
Generally ${\rm Re} \chi_0(k)$ has a deep minimum at $\omega_{\rm cont}({\bf k})$ 
that can even diverge if ${\rm Im} \chi_0(k)$ varies discontinuously. Hence, the 
charge susceptibility can develop a pole in the vicinity of the upper edge, 
which results in the onset of the ZS mode even for a small coupling. In this 
regard the $(1,0)$ and $(0,1)$-direction are special. The particle-hole 
susceptibility $\chi_0$ takes the form of a 1D-response for $\bf k$ along 
$\Gamma$-X. On a large range of doping around half filling, this results in a 
square-root singularity at $\omega_{\rm cont}({\bf k})$ which ensures the 
existence of the ZS mode along the symmetry axis and around the ${\bf k}$-point 
X. Note however that the mode is suppressed just below the UHB mode energy because 
${\rm Re}f^s(k)$ becomes negative (see Fig.~\ref{fig:fs}).

Close to half filling the ZS mode exists for nearly all momenta. As shown in 
Figs.~\ref{fig:chi_c-d_0.1} and \ref{fig:chi_c-U_8}, at strong coupling, the intensity 
of the charge response is largely transfered from the single-particle processes 
to the collective modes. Increasing the doping results in the softening of the ZS 
mode, while the response continuum grows as the quasiparticle mass is less 
renormalized. Eventually, at large doping, the ZS pole is suppressed for nearly 
all wavevectors as the singularity of ${\rm Re}\chi_0(k)$ at the continuum boundary 
is smoothed out.

\begin{figure}[b]
  \includegraphics[clip=true, width=0.48\textwidth]{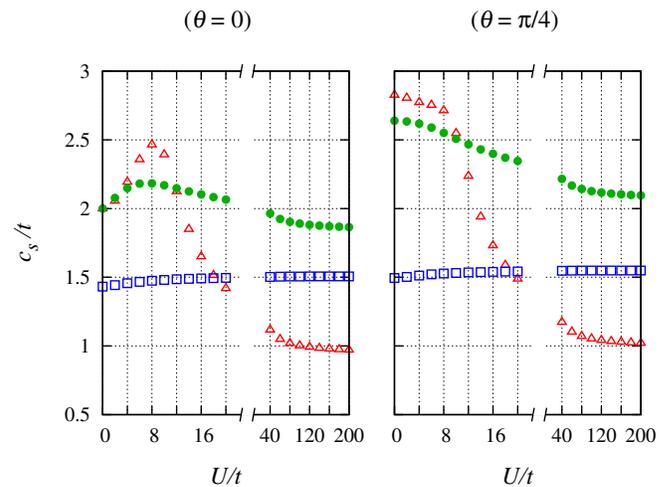}
  \caption{(Color online) Zero-sound velocity  at zero temperature as a function of 
  the coupling in the X and M-directions, for doping $\delta=0.1$ (triangle), 0.5 (dot), 
  and 0.9 (square). }
  \label{fig:zsvelocity}
\end{figure}

At long wavelength, that is in the vicinity of $\Gamma$, the dispersion of the 
pole is proportional to $|{\bf k}|$ and one can define the ZS velocity as 
\begin{equation}
c_s(\theta_{\bf k}) = \frac{\omega_{\rm ZS}({\bf k})}{|{\bf k}|}. 
\end{equation}
For the Hubbard model on the square lattice the sound velocity is anisotropic. The 
maximum is along the M-direction while the minimum is along the X-direction (see 
Fig.~\ref{fig:zsspectrum}). However the anisotropy vanishes in two limiting cases: 
at large doping $|\delta| \approx 1$ as the quasiparticle dispersion around the 
Fermi energy tends to a parabolic dispersion and, more surprisingly, close to half 
filling for strong coupling. In the latter case, the isotropy is approached because 
the ZS pole is located far above the strongly renormalized edge
$\omega_{\rm cont}({\bf k})$, at an energy where the functions $\chi_m(k)$ at long 
wavelength are dominated by their $s$-wave component.

\begin{figure}[h]
  \includegraphics[clip=true, width=0.48\textwidth]{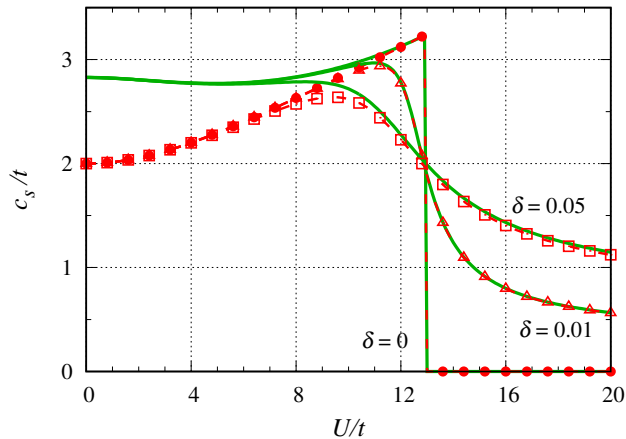}
  \caption{(Color online) Evolution of the zero-sound velocity around the metal-insulator 
  transition, in the X (symbols) and M-directions (solid lines). }
  \label{fig:zsvelocity-2}
\end{figure}

The sound velocity along the two high-symmetry directions is plotted in 
Fig.~\ref{fig:zsvelocity} for different values of coupling and doping. At small 
coupling the collective mode appears close to the continuum upper boundary which 
is $\omega_{\rm cont}({\bf k}) = z_0^2 {\rm max}_{{\bf v}_{\rm F}^0} 
({\bf v}_{\rm F}^0 \cdot {\bf k}) $ for small momentum. Here 
${\bf v}_{\rm F}^0 = \left.  \frac{ \partial t_{\bf q}}{ \partial {\bf q}} 
\right|_{{\bf q}_{\rm F}}$ is the bare Fermi velocity. Hence the weak-coupling 
approximation yields the velocity in the M-direction  
 $c_s(\frac{\pi}{4}) \approx v_{\rm F}^0 (\frac{\pi}{4}) \big[1+ \big(\frac{1}{(18 t)^2} 
 - \frac{1-\delta^2}{U_0^2}\big)U^2 \big]$
where $ v_{\rm F}^0 (\frac{\pi}{4}) = 2t\sqrt{2\big(1-({\mu}/{4t})^2\big)}$. In the 
X-direction, for doping $|\delta| \gtrsim 0.63$, the expression remains the same though 
with $v_{\rm F}^0(0) = 2t \sqrt{\big(1-(1-{|\mu|}/{2t})^2\big)}$. Otherwise for smaller 
doping we find $c_s(0) \approx 2 t \big[1+ \big(\frac{1}{(9t)^2} - \frac{1-\delta^2}{U_0^2}
\big)U^2 \big]$. The evolution of the velocity with the coupling is complicated since it 
is governed by two opposite trends. On the one hand the increase of the quasiparticle mass 
reduces it. On the other hand the increase of $f^s$ with $U$ moves the ZS pole to higher 
energy. As a result, at large doping $|\delta| \approx 1$ where the mass renormalization 
can be neglected, the velocity increases with increasing coupling. Then at a smaller 
doping the renormalization is more important and the velocity variation depends on the 
propagation angle: $c_s(\frac{\pi}{4})$ decreases while $c_s(0)$ increases before 
eventually decreasing at strong coupling. Lastly, in the vicinity of half filling the 
variation of $c_s$ is non-monotonic (see Fig.~\ref{fig:zsvelocity-2}). The velocity 
reaches a maximum at a coupling below $U_c$ before collapsing to 
$c_s \sim 2 |\varepsilon_0|\sqrt{|\delta|\big(1+\frac{U_0}{U}\big)}$ in the bad-metal 
state. The behavior at $\delta=0$ is even discontinuous: $c_s$ abruptly falls at $U_c$ 
from its maximum value $\approx 3.2t$ to zero. As previously noted the velocity around 
half filling becomes isotropic at large coupling.

\subsection{Upper-Hubbard-band mode}

A charge excitation with an energy of the order of $U$ has been predicted as 
the result of strong correlation effects since the early days of the Hubbard 
model. Indeed in the vanishing hopping limit $t=0$, all particles rest 
localized at the energy of the atomic levels $\omega=0$ or $\omega=U$. 
A perturbative inclusion of the hopping, as done by Hubbard and extended by Pairault
\textit{et al.}~\cite{Pai00}, results in the broadening of the atomic levels and the 
formation of dispersive bands around each one, the lower and upper Hubbard band. 
Hence excitations resting on the UHB are expected from this physical picture.

\begin{figure}[b]
  \includegraphics[clip=true, width=0.49\textwidth]{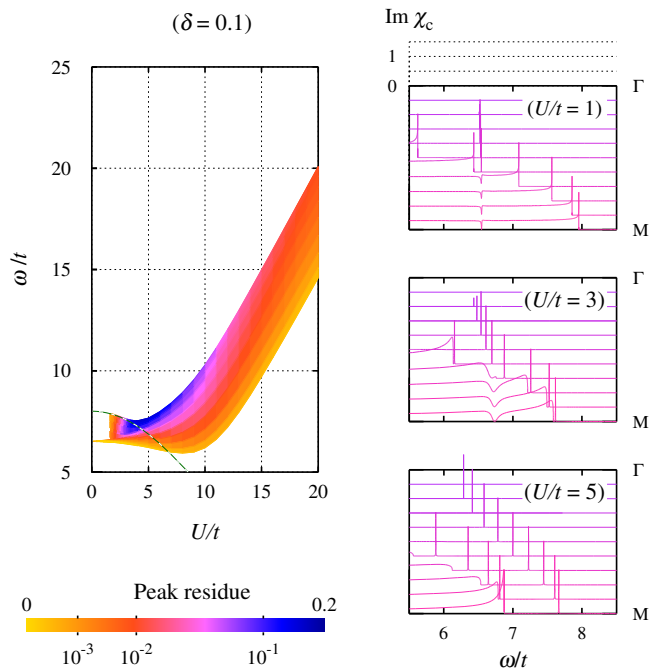}
  \caption{(Color online) Left: Coupling dependence of the UHB mode dispersion for 
  momenta along the $\Gamma$-M path, at temperature $T=t/100$ and doping $\delta=0.1$; 
  the dashed line shows the maximum energy of the response continuum, which is reached 
  at ${\rm M}$. Right: Imaginary part of $\chi_c(k)$ at coupling $U/t=1$, 3 and 5.}
  \label{fig:UHB-width-1}
\end{figure}

\begin{figure}[b]
  \includegraphics[clip=true, width=0.5\textwidth]{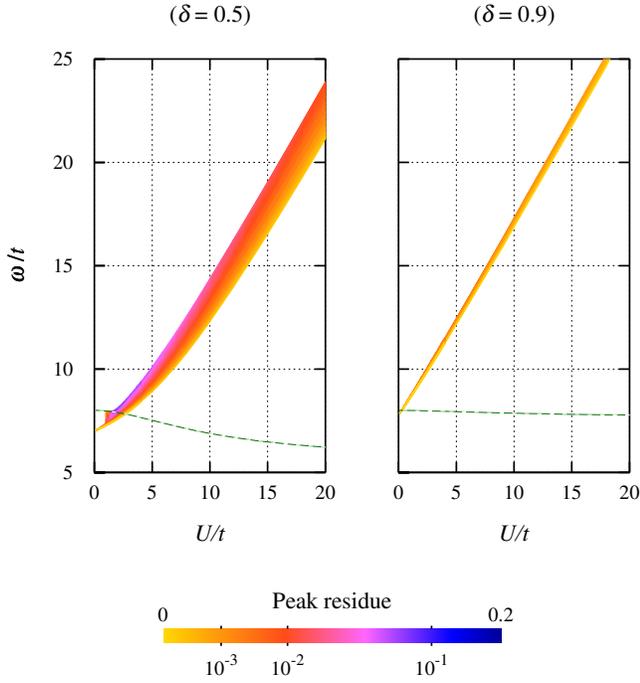}
  \caption{(Color online) Dispersion of the UHB mode for momenta along the 
  $\Gamma$-M path at temperature $T=t/100$, and doping $\delta=0.5$ and 0.9. 
  The dashed line shows the upper edge of the response continuum at ${\rm M}$.}
  \label{fig:UHB-width-2}
\end{figure}

The slave-boson approach yields such a collective excitation, below denoted 
the UHB mode, which occurs at an energy $\omega_{\rm UHB}({\bf k})$ that
grows as $U$ for strong coupling. 
Like the ZS mode, the UHB mode has an
energy dispersion with a minimum at $\Gamma$ and a maximum at M, but pushed 
to a higher energy (see Figs.~\ref{fig:chi_c-d_0.1} and \ref{fig:chi_c-U_8}). 
Actually the excitation energy does not vanish at $\Gamma$, even at small 
coupling. Numerical evaluations find that the peak weight is zero at $\Gamma$ 
and maximum at M. These features are illustrated in Figs.~\ref{fig:UHB-width-1} 
and \ref{fig:UHB-width-2} where the dispersion with momenta along 
$\Gamma-{\rm M}$ is plotted for different dopings and couplings. The mode 
appears at weak coupling around $\omega=U_0/2$ which is the frequency where 
$f^s(k)$ diverges (see the second-order expansion~(\ref{eq:f0s})). No 
dispersion is observed at the onset of the mode. Although its pole exists at 
any finite coupling, the mode disappears in the uncorrelated limit as its 
residue vanishes at $U=0$. A shift to higher energy can be observed with 
increasing doping or coupling. Their influences on the dispersion width 
are opposite. A widening is obtained by increasing the coupling while the 
effect of doping is to narrow the dispersion to the point that it vanishes 
at $|\delta| = 1$. On the whole the mode has its maximum weight at M and it is 
most clearly observed for a moderately large coupling $U \sim 5t$ at small but 
finite doping $|\delta| \sim 0.1$. Indeed its weight decreases in the close 
vicinity of half filling, and vanishes at $\delta=0$. We found that it 
is also vanishingly small at $|\delta|=1$.

The UHB mode can be distinguished from the ZS mode and the response 
continuum because its energy is generally larger than $\omega_{\rm ZS}({\bf k})$. 
However this is not necessarily the case at weak coupling. Spectra of 
${\rm Im}\chi_c(k)$ in Fig.~\ref{fig:UHB-width-1} show that it enters the 
quasiparticle continuum for momenta around M. This results in the damping of 
the excitation by quasiparticle scattering and the mode peak is replaced by a 
depletion around $\omega \approx U_0/2$ in the charge response continuum. At 
moderate coupling ($U \sim 3t$) the ZS mode that appears just beyond the 
continuum edge hybridizes with the UHB mode around M, and there is only one single 
peak around M that continuously becomes the UHB peak as ${\bf k}$ goes to 
$\Gamma$. The depletion associated to the UHB mode moves to higher energy 
with increasing coupling. After it exits the continuum the ZS mode can extend 
until M where it forms a second well-defined peak below the UHB one.

Analytical expressions for the dispersion of the UHB mode can be obtained at weak 
and strong coupling. The mode mostly occurs far beyond the continuum where 
$\chi_m(k) \sim 1/\omega^2$ and in particular 
$\chi_0(k) \approx 2 z_0^2 (\varepsilon_0 - \varepsilon_{\bf k})/\omega^2$ with
\begin{equation}
 \varepsilon_{\bf k} = \frac{2}{L} \sum_{{\bf q}} t_{{\bf q}+{\bf k}} n_F(E_{{\bf q}}). 
 \label{eq:epsk}
\end{equation} 
To first order in the high-energy expansion the denominator (\ref{eq:chi_c}) behaves 
as $\omega^2 - \omega_{\rm HB}^2$. The charge response function then possesses two 
poles, one at negative energy $\omega_{\rm LHB} = -\omega_{\rm HB}$ and one at 
positive energy $\omega_{\rm UHB} = \omega_{\rm HB}$.

At small coupling $U\ll U_0$ , the saddle-point solution can be approximated with
$x^2\approx 1 - (1-\delta^2) U/U_0$ which yields
\begin{equation}
 \omega_{\rm UHB}({\bf k}) \approx \frac{U_0}{2} \sqrt{1 + \frac{U}{2U_0} 
 \left( 1+7\delta^2 - (1-\delta^2)\frac{\varepsilon_{\bf k}}{\varepsilon_0}\right)}.
 \label{eq:wuhb1}
\end{equation}
The weak-coupling expression highlights several features of the UHB mode dispersion. 
Firstly, the collective mode appears around the energy $U_0/2$ with a dispersion that 
is vanishingly small. The expression also shows that doping results into a narrower 
dispersion that is shifted to a higher energy, as seen in Figs.~\ref{fig:UHB-width-1} 
and \ref{fig:UHB-width-2}. The dispersion width is approximately equal to 
$(1-\delta^2) U/4$ and it vanishes for $|\delta|=1$. 

The approximation in the strong-coupling limit is obtained with
$x^2 \approx |\delta|/{\sqrt{1- \frac{U_0}{U} \Big( 1- {|\delta|}/{\sqrt{1-U_0/U}} \Big)}}$ 
which gives
\begin{equation}
 \omega_{\rm UHB}({\bf k}) \approx U \sqrt{1 - \frac{U_0}{2U} \left( 1-3|\delta| 
 + (1-|\delta|) \frac{\varepsilon_{\bf k}}{\varepsilon_0}\right)}. 
  \label{eq:wuhb2}
\end{equation}
The dispersion thus has its minimum $\approx U - U_0(\frac{1}{2} - |\delta|)$ at 
$\Gamma$, and its width is approximately $(1-|\delta|) U_0/2$. Hence at large 
coupling the energy of the mode grows linearly as the on-site Coulomb interaction $U$. 
This genuine strong correlation effect is one of the most important results of this 
work. Being of order $U$ the mode follows from the UHB, that is not captured by the 
conventional HF~+~RPA approach. It should also be emphasized that 
Eqs.~(\ref{eq:wuhb1}) and (\ref{eq:wuhb2}) hold for arbitrary lattices with one atom 
in the unit cell, irrespective of the dimensionality. 

\subsection{Effect of temperature}

\begin{figure}[b]
 \includegraphics[clip=true, width=0.49\textwidth]{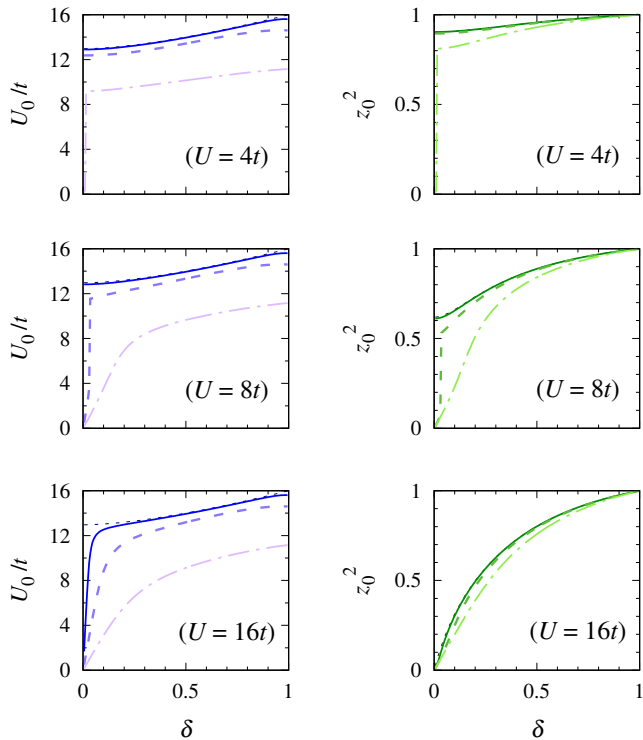}
  \caption{(Color online) Doping dependence of the coupling scale $U_0$ and the 
  inverse-mass renormalization factor $z_0^2$ at temperature $T=0$ (dotted line), 
  $t/10$ (solid line), $t/3$ (dashed line), and $t$ (dot-dashed line), for different 
  values of on-site interaction $U$.}
  \label{fig:U0}
\end{figure}

Within our theory the impact of temperature on the collective modes manifests itself 
in two different ways.
Firstly the collective mode dispersion shrinks with increasing temperature. This 
results from the decrease of the saddle-point values, most notably for doping 
$|\delta| \lesssim 0.1$ and strong coupling. As shown in Fig.~\ref{fig:eps0_U0} 
and Fig.~\ref{fig:U0} the averaged kinetic energy $\varepsilon_0$, the coupling scale 
$U_0$, and the inverse-mass renormalization factor $z_0^2$ vary significantly with 
temperature for this regime of parameters. However in this region of the phase 
diagram (Fig.~\ref{fig:phasediagram}) the paramagnetic solution is unstable toward 
phase separation or incommensurate magnetic ordering. Outside this regime, where our 
investigation is of better relevance, the effect of temperature is a mild reduction 
of the amplitudes of the saddle-point values. Thus increasing the temperature 
up to $T = t/3$ slightly scales down the spectrum along the energy axis. We 
will not discuss the regime of high temperature where the approximation of Gaussian 
fluctuations certainly becomes insufficient. For instance we expect that incoherent
multi-particle processes, which are not taken into account here, get more 
prominent and modify significantly the charge response of the system, as exemplified 
by the physics of liquid helium.

The second notable effect of temperature is the broadening of the collective mode 
peak, which results from scattering of thermally excited quasiparticles. Let us 
first remark that in the absence of incoherent multi-particle processes, the peak is not 
damped above the energy $\Delta E_{\rm max}({\bf k}) = 4 t z_0^2 (|\sin 
\frac{k_x}{2}| + |\sin \frac{k_y}{2}|)$ of the most energetic one-particle transition 
with momentum transfer $\bf k$. The UHB peak generally lies above it so its shape 
is hardly affected by increasing the temperature. This is not the case of the ZS mode 
for small wavevectors at large doping, and in the vicinity of $\Gamma$ at any finite 
doping. The ZS peak is broadened because the charge response continuum does not 
extend up to $\Delta E_{\rm max}({\bf k})$ for small wavevectors. The reason comes 
from the Fermi statistics which, at zero temperature, excludes some one-particle 
transitions, among which can be found the most energetic one that occurs between the 
states of momenta $(\frac{\pi -k_x}{2},\frac{\pi -k_y}{2})$ and $(\frac{\pi+ k_x}{2},
\frac{\pi + k_y}{2})$. As a result the ZS peak can be located between the continuum 
upper edge $\omega^{T=0}_{\rm cont}({\bf k})$ at $T=0$ and $\Delta E_{\rm max}
({\bf k})$. Increasing the temperature then smears the Fermi distribution, which 
populates the response continuum in this energy range, and eventually broadens 
the ZS peak.

\section{Comparison with other approaches}
\label{sec:comparison} 

\subsection{Comparison with HF~+~RPA result}

\begin{figure*}
  \includegraphics[clip=true, width=0.95 \textwidth]
  {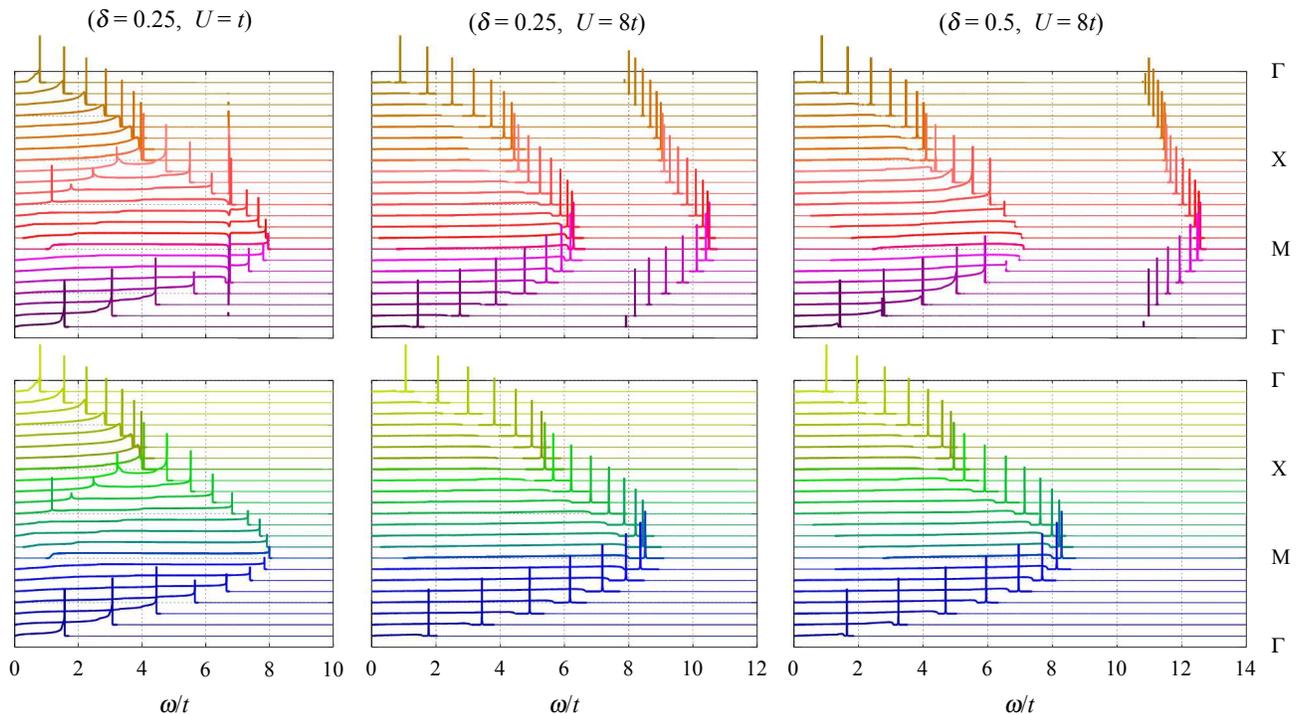}
  \caption{(Color online) Comparison between the imaginary part of $\chi_c(k)$ (top) 
  and of $\chi_{\rm RPA}(k)$ (bottom) at moderate coupling $U=t$ and strong
  coupling $U=8t$. Parameter: $T=t/100$.}
  \label{fig:RPA_compare}
\end{figure*}

In the weak-coupling limit the charge response obtained within the slave-boson 
method is mostly similar to the standard HF~+~RPA result. But, as stated earlier, 
the former possesses a supplementary collective mode at high energy, the UHB mode. 
And, although the perturbation method also produces a ZS mode, it fails to account 
for the correlation effects which strongly renormalize the quasiparticle mass 
around half filling and for the dynamical screening of the electron interaction. 
This is shown in Fig.~\ref{fig:RPA_compare} where the slave-boson charge response 
is compared with the HF~+~RPA response 
\begin{equation}
 \chi_{\rm RPA}(k) = \frac{\chi_0^{(0)}(k)}{1 + \frac{U}{2} \chi_0^{(0)}(k)}.
\end{equation}
Here $\chi_0^{(0)}(k)$ is the charge response function of a Fermi gas, {\it i.e.} with 
no mass renormalization. At moderate coupling $U=t$ the only observable difference
between the two responses is the dispersionless UHB mode. The contribution of the 
latter is small and the weight of its peak actually vanishes in the limit $U=0$.
However at large coupling $U=8t$, the two spectra are quite different. The 
slave-boson response has two well separated collective modes while the perturbation 
method only yields the ZS mode. Furthermore, the continuum width and the ZS 
dispersion shrink due to the quasiparticle mass enhancement, whereas such a
correlation effect is not captured by $\chi_{\rm RPA}(k)$. The mass renormalization
is not the only effect of correlations. At large doping, the ZS peak in $\chi_c(k)$
disappears around ${\bf k}={\rm M}$, in contrast to the HF~+~RPA prediction. This is
because the bare electron interaction $U/2$ of the perturbation result is replaced by 
the complex function $f^s(k)$ within the slave-boson approach. The latter depends on 
frequency and momentum, and it can have a negative real part near 
$\omega_{\rm cont}({\bf k})$ (see Fig.~\ref{fig:fs}) which thus suppresses the ZS pole.

\subsection{Comparison with time-dependent GA}

The Kotliar and Ruckenstein slave-boson approach has historically been designed 
to reproduce the Gutzwiller approximation at the saddle-point level~\cite{Kot86}, 
thereby strongly tiding both schemes. Later on, a method to calculate excitations 
at zero temperature has been built on the GA and the RPA~\cite{Sei01}. It takes 
the form of the above RPA series with an effective interaction, therefore missing 
the physics of the Hubbard split bands in the charge response function. Yet a 
refined treatment has been proposed in Ref.~\cite{Bun13}, which we now compare to 
the slave-boson result. 

We restrict the analysis to the double-occupancy excitations for which the 
comparison is simplified. We note that all three terms in Eq.~(\ref{eq:chi_c-def}) 
contribute to the particle-hole continuum, implying a damping in the 
double-occupancy excitation spectra that is absent from the time-dependent 
Gutzwiller approximation (TDGA)~\cite{Bun13}. From a quantitative point of view, 
one can observe that the pole of the double-occupancy propagator found by the TDGA 
Eq.~(100) in \cite{Bun13} is located at an energy smaller that the slave-boson one. 
The discrepancy is largest at ${\bf k}={\rm M}$, for small doping, and strong 
coupling. For instance the TDGA (slave-boson) pole disperses from $\omega/t=7.1$ 
to 8.8 (7.2 to 10) for $U=8t$ and $\delta=0.2$, and from $\omega/t=14.4$ to 17.8 
(14.6 to 20.1) for $U=20t$ and $\delta=0.1$. Hence the excitations computed within
TDGA show both qualitative and quantitative differences to our results which are
controlled by the $1/N$ expansion~\cite{FW}.

\subsection{Comparison with numerical methods}

We have compared the charge response function evaluated with the slave-boson method 
to exact diagonalization (ED) data~\cite{Bun13,Jia12,Wan14} and quantum Monte Carlo 
(QMC) simulations~\cite{Zim97,DziUnpub,Che94,Buh99,Bec00,Pre97,Gro00,Koh04,Kun15}
available in the literature. The low-temperature phase found by the numerical 
methods at half filling is an antiferromagnetic insulator. 
But, as confirmed by our investigation of instabilities, the paramagnetic  
solution becomes predominant with increasing doping and temperature. 
Keeping this in mind when comparing our evaluation of the charge response, we note 
that the spectra computed at finite doping by both numerical approaches show salient 
features that can be naturally explained by the two collective modes found in the 
present work. In particular the variations of their dispersions with the coupling 
and the doping qualitatively agree with the behavior we have described.

ED are performed on finite clusters and the small size of the system enhances the
energy level separation. As a result the obtained spectrum is a set of peaks rather 
than a continuous function of the frequency. The energy quantization is visible
in the spectra of the charge susceptibility calculated in Ref.~\cite{Bun13} at small 
density $\langle \cali{N} \rangle \approx 0.03$. They remarkably show 
two distinctive peaks at the energies where we have found the ZS peak at 
$\bf{k}=$X and the UHB peak. Confirming our results, the first peak is 
exactly in the middle of the main contribution to the momentum-integrated response, 
that corresponds to the continuum of single-particle excitations. As for the 
second peak at higher energy, we note that the dispersion of the ED computation is narrow 
and the peak weight is vanishingly small, which can be explained by the UHB mode found
by our theory close to doping $|\delta|= 1$. The charge response function has also 
been computed around half filling, but for the Hubbard model including hopping between 
next-nearest-neighbor sites~\cite{Jia12,Wan14}. The latter is known to break 
particle-hole symmetry. So the comparison with our results for the simple Hubbard model 
should be taken with caution. One can nevertheless remark a encouraging agreement for 
hole doping. For the large value of coupling $U=10t$ the ZS-like structure at the 
boundary of the continuum is found to decrease in energy with increasing hole doping, 
which is also predicted by the slave-boson method. Besides the high-energy feature 
moves up in energy as the UHB mode.

Early QMC simulations of the Hubbard model have been mainly focused on the static
spin and charge structure factors~\cite{Che94,Buh99,Bec00}. As previously discussed 
in Ref.~\cite{Zim97}, the SRI slave-boson approach is in a very good quantitative 
agreement with the numerical evaluations of these quantities. Concerning the dynamical
response functions, and in particular the search for collective modes, the analysis of 
the QMC results meets two hurdles. Indeed statistical averages computed by QMC 
simulations yield the values of the correlation functions on the imaginary-frequency 
axis. Their values on the real-frequency axis are then approximated by different 
numerical schemes, such as the maximum entropy method, which limits the obtained 
frequency definition. Furthermore the simulations of a doped system are restricted to 
the high-temperature regime $T\gtrsim t/3$ by the sign problem. As a consequence QMC 
spectra may lack the necessary energy resolution to distinguish fine structures, such 
as several collective-mode peaks close to one another, or a peak with a small weight 
which is the case of the UHB mode for a large set of parameters. 

The charge response function of a doped system is computed in the QMC 
simulations~\cite{Pre97,Gro00,Koh04,Kun15} at temperature $T=t/3$ for coupling 
values $U=4t$ and $8t$. At this temperature a sensible comparison with our 
theory may be made for doping $|\delta| \gtrsim 0.1$ at which the 
paramagnetic phase should prevail. The slave-boson results are consistent 
with the obtained QMC spectra. The latter show that the continuum 
response is strongly reduced at low doping and the intensity is mainly located 
beyond it, around the energies of the ZS and the UHB modes. For instance for 
${\bf k}= M$ and $U=8t$, the intensity mainly spreads from $\omega \approx 4t$ 
to $\omega \approx 12t$. This can be interpreted as the response of the collective 
modes that interact with the background of incoherent multi-particle processes. 
By increasing the doping, the UHB mode energy increases and because of its small 
weight, its signature can no longer be distinguished from the structureless 
background in the QMC spectra. Meanwhile the continuum response is less 
renormalized away from half filling and the ZS mode energy decreases. The most 
satisfying comparison is found with the QMC simulations of Ref.~\cite{Koh04} 
performed for $U=4t$. The spectra show two clear structures, one similar to the 
ZS peak at the edge of the continuum response, and the other one around 
$\omega \approx 8t$ which possesses a slight dispersion as the UHB mode. 

\section{Conclusion} \label{sec:conclusion}

We have derived the expression of the charge susceptibility
of the Hubbard model in its Kotliar and Ruckenstein slave-boson
representation. We have shown that it reduces to the conventional HF~+~RPA
result when expanded to lowest order in $U$. They markedly depart from
one another already to next order in $U$. We then investigated spin and
charge instabilities as well as charge collective modes of the 2D
Hubbard model in the thermodynamical limit. To that aim we used the spin
rotation invariant formulation of the above representation. Extending
previous work, our calculations showed that magnetic instabilities of
the paramagnetic phase essentially disappear for temperature $ T \geq
t/6$, which lays ground for the computation of the charge susceptibility
in this regime. In the strong coupling regime, the charge excitation
spectrum splits into a low frequency branch, and a high frequency
collective mode. \textit{En passant}, an approximated analytical form of
the latter has been derived. It applies to arbitrary lattices containing
one site in the unit cell. This mode, that may not be accounted for
within the conventional HF~+~RPA framework or self-consistent perturbative
schemes such as FLEX, disperses around $\omega \simeq U$ and therefore
follows from the upper Hubbard band.

At low energy the charge excitations form a continuum, which width
scales with the quasiparticle residue $z_0^2$, again in contrast to the
conventional HF~+~RPA framework result. A collective mode lies above its
upper boundary. The velocity of this zero-sound mode is anisotropic both
off half-filling and away from the low density limit. We did not find a
universal behavior in its dependence on the coupling strength, because
it results from two opposite trends: on the one hand the increase of the
effective mass reduces it while on the other hand the zero-sound
excitation is shifted to higher energy. Nevertheless some trends could
be identified; for instance it shows very small dependence on $U$ in the
small density regime. Furthermore, for small to intermediate doping, the
zero-sound velocity decreases, once $U$ exceeds the band width. To some
extend, our results could be interpreted within Pines' polarization potential 
theory. Indeed striking similarities are found at low frequency when the ZS 
and UHB modes are well split. Yet the polarization potential theory does not 
entail a UHB mode and it therefore fails to describe the regime where the ZS 
and the UHB modes strongly hybridize. We also studied the temperature dependence of the 
charge excitation spectrum. We found the small wavevector zero-sound 
excitation to broaden with increasing temperature, while the other features 
show little temperature dependence. 

\section*{Acknowledgments}

We gratefully thank D. Braak, T. Kopp, M. Raczkowski, and A.-M. S. Tremblay 
for several stimulating discussions. The authors acknowledge the financial 
support of the French Agence Nationale de la Recherche (ANR), through
the program Investissements d'Avenir (ANR-10-LABX-09-01), LabEx EMC3, the
R\'egion Basse-Normandie, the R\'egion Normandie, and the Minist\`ere de la
Recherche.

\appendix

\section{Elements of the fluctuation matrix $S_{ij}$}
\label{app_S}

The fluctuation matrix is symmetric except for off-diagonal elements 
$S_{\mu3}(k) = - S_{3\mu}(k)$. It is composed of four blocks, one for the charge 
fluctuations and three for the spin fluctuations.

As emphasized by~\cite{Kot92,FW}, it is essential to notice the absence 
of a full radial gauge in order to describe the UHB mode. Indeed, in the 
early calculations~\cite{Ras88,Lav90,Lil90,li94} the erroneous conclusion that 
the phase of all slave-boson fields could be gauged away resulted in a 
$5 \times 5$ matrix for the fluctuation matrix in the charge channel. 
However following the observation that one slave-boson field has to be 
complex yields a $6 \times 6$ matrix that possesses the supplementary 
dynamics introduced by the time derivative of this boson field~\cite{Jol91}.
As a result the charge susceptibility acquires an $\omega^2$-dependence in
addition to the frequency dependence contained in the fermionic bubbles 
$\chi_m(k)$, and a second pole describing the UHB mode. The charge 
fluctuation matrix has thus been obtained in the limit ${\bf q}=0$ and 
$\delta=0$ where the softening of the UHB mode has been found at the 
Mott-Hubbard transition~\cite{Kot92}. Later a general expression of 
$\chi_c(k)$ for arbitrary momentum and density has been derived within 
the SRI representation~\cite{Zim97}. But, as stated earlier, it does not 
include several matrix elements that are present in the correct 
expression~(\ref{eq:chi_c}). It turns out that the missing terms do not 
contribute to the correlation functions in the limits $\omega=0$ or 
${\bf q}=0$, which may explain why they have been overlooked until now. 
However they are crucial to reproduce the RPA result at weak coupling.

The non-zero terms of the charge part are
\begin{equation}
S_{1,1}(k) = \alpha + s_{1,1}(k) \nonumber 
\end{equation}

\begin{equation}
S_{\mu\nu}(k)  = s_{\mu\nu}(k) \;\;  \text{for $\mu,\nu=1,2,4$ with $\mu\neq \nu$} 
\nonumber 
\end{equation}

\begin{equation}
S_{\mu 3}(k)  = -S_{3\mu}(k) = - \frac{{\rm i} \nu_n}{2}\chi_1(k) 
\frac{\partial z}{\partial \psi_{\mu}} \frac{\partial z}{\partial d''}  
\;\; \text{for $\mu=1,4$} \nonumber 
\end{equation}

\begin{equation}
S_{1,5}(k) = - \frac{1}{2} \chi_1(k) z_0 \frac{\partial z}{\partial e} \nonumber 
\end{equation}  

\begin{equation}
S_{1,6}(k) = e \nonumber 
\end{equation}

\begin{equation}
S_{2,2}(k) = \alpha - 2\beta_0 + U + s_{2,2}(k) \nonumber 
\end{equation}
  
\begin{equation}
S_{2,3}(k) = - S_{3,2}(k) = \nu_n \left(1  - \frac{\rm i}{2} \chi_1(k) 
\frac{\partial z}{\partial d'} \frac{\partial z}{\partial d''} \right)\nonumber 
\end{equation}  
  
\begin{equation}
S_{2,5}(k) = -2d - \frac{1}{2} \chi_1(k) z_0 \frac{\partial z}{\partial d'} \nonumber 
\end{equation} 

\begin{equation}
S_{2,6}(k) = d \nonumber 
\end{equation}  
  
\begin{equation}
S_{3,3}(k) = \alpha - 2 \beta_0 + U + s'_{3,3}(k) \nonumber 
\end{equation}

\begin{equation}
S_{3,5}(k)  = -S_{3,5}(k) = - \frac{{\rm i} \nu_n}{2 z_0} \chi_0(k)  
\frac{\partial z^*}{\partial d''}  \nonumber 
\end{equation}

\begin{equation}
S_{4,4}(k) = \alpha - \beta_0 + s_{4,4}(k) \nonumber 
\end{equation}

\begin{equation}
S_{4,5}(k) = -p_0 - \frac{1}{2} \chi_1(k) z_0 \frac{\partial z}{\partial p_0} \nonumber 
\end{equation} 

\begin{equation}
S_{4,6}(k) = p_0 \nonumber 
\end{equation}

\begin{equation}
S_{5,5}(k) = - \frac{1}{2} \chi_0(k) 
\end{equation} 
The spin blocks are given by
\begin{equation}
 S_{7,7}(k) = S_{9,9}(k) = S_{11,11}(k) = \alpha - \beta_0 + s_{11,11}(k) \nonumber
\end{equation}
\begin{equation}
 S_{8,8}(k) = S_{10,10}(k) = S_{12,12}(k) = -\frac{1}{2} \chi_0(k) \nonumber
\end{equation}
\begin{equation}
 S_{7,8}(k) = S_{9,10}(k) = S_{11,12}(k) = -p_0 -\frac{1}{2} \chi_1(k) 
 \frac{\partial z_{\uparrow} }{\partial p_3} z_0 
\end{equation}
We have used
\begin{equation}
 s_{\mu\nu}(k) = \varepsilon_0 z_0 \frac{\partial^2 z}{\partial \psi_{\mu} 
 \partial \psi_{\nu}} + \left[ \varepsilon_{{\bf k}} - \frac{1}{2} z_0^2 \chi_2(k) \right] 
 \frac{\partial z}{\partial \psi_{\mu}} \frac{\partial z}{\partial \psi_{\nu}} 
\end{equation}

\begin{equation}
 s'_{3,3}(k) = \varepsilon_0 z_0 \frac{\partial^2 z}{\partial d''\partial d''} 
 + \left[ \varepsilon_{0} + \frac{\nu_n^2}{2 z_0^2} \chi_0(k) \right] 
 \left|\frac{\partial z}{\partial d''} \right|^2 
\end{equation}
with the fermionic bubbles
\begin{equation}
 \chi_m(k) = \frac{2}{L} \sum_{{\bf q }} (t_{{\bf q }+{\bf k}} + t_{{\bf q }})^m 
 \frac{n_F(E_{{\bf q }+{\bf k}}) - n_F(E_{{\bf q }})}
 {\ii \nu_n - (E_{{\bf q }+{\bf k}} - E_{{\bf q }}) }
\end{equation}
and $\varepsilon_{\bf k}$ is given by Eq.~(\ref{eq:epsk}).

The expressions of the derivatives of $z$ may be gathered from Ref.~\cite{li91,Zim97}. 
Note however a misprint in \cite{Zim97} which should be corrected as
\begin{equation}
 \frac{\partial^2z}{\partial {d'}^2} = \frac{2\sqrt{2} p_0 \eta}{1+\delta} \left(2d + x 
 + \frac{6 x d^2}{1+\delta} \right)
\end{equation}

\section{Spin susceptibility $\chi_s(k)$}
\label{app_chi_s}

As shown by~\cite{Lav90,li91,Zim97}, inverting the fluctuation matrix $S$ yields the spin dynamical response function
\begin{equation}
 \chi_s(k) = \frac{\chi_0(k)}{1 + A_{\bf k} \chi_0(k) + B \chi_1(k) + C \big( \chi_1^2(k) 
 - \chi_0(k) \chi_2(k) \big)}
 \label{eq:chi_s}
\end{equation}
where

\begin{align}
 A_{\bf k} & = \frac{1}{2 p_0^2} \left[ \alpha - \beta_0 + \varepsilon_0 z_0 
 \frac{\partial^2 z_{\uparrow}}{\partial p_3^2} + \varepsilon_{\bf k} 
 \left( \frac{\partial z_{\uparrow}}{\partial p_3}\right)^2 \right], \nonumber \\
 B & = \frac{z_0}{p_0} \frac{\partial z_{\uparrow}}{\partial p_3}, \nonumber \\
 C & = \left( \frac{z_0}{2 p_0}\right)^2 \left( \frac{\partial z_{\uparrow}}
 {\partial p_3}\right)^2. 
\end{align}

Similarly to the charge dynamical response function, we have found that in the 
weak-coupling limit the expression can be simplified as
\begin{equation}
 \chi_s(k)= \frac{\chi_0(k)}{1 + f^a(k) \chi_0(k)}
\end{equation}
where $f^a(k)$ can be expanded to the second order in $U$ as
\begin{equation}
 f^a(k) = \frac{U}{2} \left[ -1 + \frac{U}{2 U_0} \bigg( 3 + 5 \delta^2 - (1-\delta^2)
 \delta \gamma(k)\bigg)\right]
\end{equation}
with $\gamma(k)=\chi_1(k)/\varepsilon_0 \chi_0(k)$. Reducing this result to
first order in $U$ yields an exact agreement with the perturbation theory, and 
the Landau's Fermi-liquid spin parameter is obtained with $F_0^a = N_F f^a(0)$.


\begin{thebibliography}{0}

\bibitem{Hub63} J. Hubbard, 
                    Proc. Roy. Soc. London A \textbf{276}, 238 (1963);
                    \textit{ibid.} \textbf{281}, 401 (1964).
                    
\bibitem{Lan56} L.~D. Landau,
                   Sov. Phys. JETP \textbf{3}, 920 (1956).
                    
\bibitem{PinBoh} D. Bohm and D. Pines, 
                    Phys. Rev. \textbf{82}, 625 (1951); 
                    \textit{ibid.} \textbf{85}, 338 (1952); 
                    \textit{ibid.} \textbf{92}, 609 (1953). 
                    
\bibitem{Edw68} D.~M. Edwards and A.~C. Hewson, Rev. Mod. Phys. \textbf{40}, 810 (1968).     

\bibitem{PinNoz} D. Pines and P. Nozi\`eres, 
                    \textit{Theory of Quantum Fluids}, Vol. 1 (Benjamin, New York, 1966).
                    
\bibitem{Vol84} D. Vollhardt, 
                    Rev. Mod. Phys. \textbf{56}, 99 (1984).
                  
\bibitem{Bor91} D. Bormann, T. Schneider, and M. Frick,
                    Europhys. Lett. \textbf{14}, 101 (1991).
 
\bibitem{Geo96} A. Georges, G. Kotliar, W. Krauth, and M. J. Rozenberg, 
                    Rev. Mod. Phys. \textbf{68}, 13 (1996).
                    
\bibitem{Bul01} R. Bulla, T. A. Costi, and D. Vollhardt,
                   Phys. Rev. B \textbf{64}, 045103 (2001).

\bibitem{Shr90} B.~I. Shraiman and E.~D. Siggia, 
                    Phys. Rev. Lett. \textbf{62}, 1564 (1989).                  

\bibitem{Fre91} R. Fr\'esard, M. Dzierzawa, and P. W\"olfle,
                    Europhys. Lett. \textbf{15}, 325 (1991).
                    
\bibitem{SeiSi} G. Seibold, E. Sigmund, and V. Hizhnyakov,
                    Phys. Rev. B \textbf{57}, 6937 (1998).    
             
\bibitem{Sei02} J. Lorenzana and G. Seibold,
                    Phys. Rev. Lett. \textbf{89}, 136401 (2002);
                                     \textbf{90}, 066404 (2003);
                                     \textbf{94}, 107006 (2005).

\bibitem{Rac06} M. Raczkowski, R.~Fr\'esard, and A.~M. Ole\'s,
                    Phys. Rev. B \textbf{73}, 174525 (2006);
                 M. Raczkowski, M. Capello, D. Poilblanc, R. Fr\'esard,
                    and A. M. Ole\'s,
                    \textit{ibid.} \textbf{76}, 140505(R) (2007).
                                        
\bibitem{Cha10} C.~C. Chang and S. Zhang, 
                    Phys. Rev. Lett. \textbf{104}, 116402 (2010).
                   

\bibitem{stripes} J. Zaanen and O. Gunnarsson, 
                      Phys. Rev. B \textbf{40}, 7391 (͑1989); 
                  D. Poilblanc and T. M. Rice, 
                      ibid. \textbf{39}, 9749 (1989); 
                  H. J. Schulz, 
                      J. Phys. France \textbf{50}, 2833 (1989); 
                      Phys. Rev. Lett. \textbf{64}, 1445 (1990);
                  K. Machida, 
                      Physica C \textbf{158}, 192 (1989); 
                  M. Inui and P. B. Littlewood,
                      Phys. Rev. B \textbf{44}, 4415 (1991);
                  S. A. Kivelson, E. Fradkin, and V. J. Emery, 
                      Nature \textbf{393}, 550  (1998).    
                      
\bibitem{Cor14}  P. Corboz, T. M. Rice, and M. Troyer,
                      Phys. Rev. Lett. \textbf{113}, 046402 (2014).
                    
\bibitem{Lep15} A. Lepr\'evost, O. Juillet, and R. Fr\'esard,
                    New J. Phys. \textbf{17}, 103023 (2015).
                 
\bibitem{Haf14} H. Hafermann, E.~G.~C.~P. van~Loon, M.~I. Katsnelson, 
                A.~I. Lichtenstein, and O. Parcollet,
                    Phys. Rev. B \textbf{90}, 235105 (2014).

\bibitem{Kot86} G. Kotliar and A. E. Ruckenstein,
                    Phys. Rev. Lett. \textbf{57}, 1362 (1986).

\bibitem{Bri70} W. F.~Brinkman and T. M.~Rice,
                    Phys. Rev. B \textbf{2}, 4302 (1970).

\bibitem{Li89}  T. C. Li, P. W\"olfle, and P. J. Hirschfeld,
                    Phys. Rev. B \textbf{40}, 6817 (1989).

\bibitem{FW}    R. Fr\'esard and P. W\"olfle,
                    Int. J. of Mod. Phys. B \textbf{6},  685 (1992);
                                            \textbf{6}, 3087 (1992).
\bibitem{Lil90} L. Lilly, A. Muramatsu, and W. Hanke,
                    Phys. Rev. Lett. \textbf{65}, 1379 (1990).

\bibitem{Igo13} P.~A. Igoshev, M.~A. Timirgazin, A.~K. Arzhnikov, and V.~Y. Irkhin, 
                    JETP Lett. \textbf{98}, 150 (2013).

\bibitem{Fre92} R. Fr\'esard and P. W\"olfle,
                    J. Phys.: Condens. Matter \textbf{4}, 3625 (1992).
                    
\bibitem{Doll2} B. M\"oller, K. Doll, and R. Fr\'esard,
                    J. Phys.: Condensed Matter \textbf{5}, 4847 (1993).


\bibitem{Fle01} M. Fleck, A. I. Lichtenstein, and A.~M. Ole\'s,
                    Phys. Rev. B \textbf{64}, 134528 (2001).

\bibitem{RaEPL} M. Raczkowski, R.~Fr\'esard, and A.~M.~Ole\'s,
                    Europhys. Lett. \textbf{76}, 128 (2006).
                    
\bibitem{Fre02} R. Fr\'esard and M. Lamboley, 
                    J. Low Temp. Phys. \textbf{126}, 1091 (2002). 
                    
\bibitem{lhoutellier15} G. Lhoutellier, R. Fr\'esard, and A. M. Ole\'s, 
                        Phys. Rev. B {\bf 91}, 224410 (2015). 
                        
\bibitem{FW98} R. Fr\'esard and W. Zimmermann, 
                   Phys. Rev. B \textbf{58}, 15288 (1998).

\bibitem{Igo15} P.A. Igoshev, M.A. Timirgazin, V.F. Gilmutdinov,
                A.K. Arzhnikov, V.Yu. Irkhin,
                   J. Phys: Condens. Matter \textbf{27}, 446002 (2015).

\bibitem{Kot00} G. Kotliar, E. Lange, and M. J. Rozenberg,
                    Phys. Rev. Lett. \textbf{84}, 5180 (2000).

\bibitem{Doll3} R.~Fr\'esard and K.~Doll, Proceedings of the NATO ARW
                    \textit{The Hubbard Model: Its Physics and
                    Mathematical Physics}, eds. D. Baeriswyl, D. K. Campbell,
                    J.~M.~P.~Carmelo, F. Guinea, and E. Louis,
                    San Sebastian (1993) (Plenum Press, 1995), p. 385.

\bibitem{Pav06} N. Pavlenko and T. Kopp,
                   Phys. Rev. Lett. \textbf{97}, 187001 (2006).

\bibitem{Ste17} K. Steffen, R. Fr\'esard, and T. Kopp, 
                   Phys. Rev. B \textbf{95}, 035143 (2017).                   

\bibitem{Zim97} W. Zimmermann, R. Fr\'esard, and P. W\"olfle,
                    Phys. Rev. B \textbf{56}, 10097 (1997).

\bibitem{fresard12} R. Fr\'esard, J. Kroha, and P. W\"olfle, 
in {\it Theoretical Methods for Strongly Correlated Systems}, 
edited by A. Avella and F. Mancini, 
Springer Series in Solid-State Sciences Vol. 171 
(Springer-Verlag, Berlin, 2012), pp. 65-101.

\bibitem{li91} T. Li, Y. S. Sun, and P. W\"olfle, 
Z. Phys. B {\bf 82}, 369 (1991).


\bibitem{Fre01} R. Fr\'esard and T. Kopp,
                   Nucl. Phys. B \textbf{594}, 769 (2001).

\bibitem{Kop07} R.~Fr\'esard, H.~Ouerdane, and T.~Kopp,
                   Nucl. Phys. B \textbf{785}, 286 (2007).

\bibitem{Jol91} Th. Jolic{\oe}ur and J. C. Le Guillou,
                   Phys. Rev. B \textbf{44}, 2403 (1991).

\bibitem{Kot92} Y. Bang, C. Castellani, M. Grilli, G. Kotliar, R. Raimondi,
                    and Z. Wang,
                    Int. J. of Mod. Phys. B \textbf{6}, 531 (1992);
                    Proceedings of the Adriatico Research Conference and
                    Miniworkshop \textit{Strongly Correlated Electrons Systems
                    III}, eds. Yu Lu, G. Baskaran, A. E. Ruckenstein, E.
                    Tossati (World Scientific Publishing Co., Singapore, 1992).

\bibitem{Met89} W.~Metzner, and D.~Vollhardt,
                    Phys. Rev. Lett. \textbf{62}, 324 (1989);
                    Phys. Rev. B \textbf{37}, 7382 (1988);
                 W.~Metzner,
                    Z.~Phys. B \textbf{77}, 253 (1989).
                    
\bibitem{Vol87} D. Vollhardt, P. W\"olfle, and P. W. Anderson,
                    Phys. Rev. B \textbf{35}, 6703 (1987).
                                        

\bibitem{FK97} R. Fr\'esard and G. Kotliar,
                   Phys. Rev. B \textbf{56}, 12 909 (1997).                                        
                                        
\bibitem{camjayi07} A. Camjayi, M. J. Rozenberg, and R. Chitra, 
                    Phys. Rev. B {\bf 76}, 195108 (2007).
                    
\bibitem{McWhan73} D.~B.~McWhan, A. Menth, J. P. Remeika, W. F. Brinkman, and T. M. Rice, 
                   Phys. Rev. B \textbf{7}, 1920 (1973).
                   
\bibitem{doll93} K. Doll, M. Dzierzawa, R. Fr\'esard, and P. W\"olfle, 
                 Z. Phys. B {\bf 90}, 297 (1993).

\bibitem{Lav90} M. Lavagna,
                    Phys. Rev. B \textbf{41}, 142 (1990).
                        
\bibitem{li94} T. Li and P. B\'enard, 
               Phys. Rev. B {\bf 50}, 17837 (1994).   
               
\bibitem{DziUnpub} M. Dzierzawa (unpublished).                    

\bibitem{Pin81} D. Pines,
                    in \textit{Quantum Fluids}, ed. D. F. Brewer
                    (North-Holland, Amsterdam, 1966), p. 257;
                C. H. Aldrich and D. Pines,
                    J. Low Temp. Phys. \textbf{32}, 689 (1978);
                for a review see D. Pines,
                    Lecture Notes in Phys. \textbf{142}, 202 (1981).  
                    
\bibitem{Pai00} S. Pairault, D. S\'en\'echal, A.-M.S. Tremblay,
                    Eur. Phys. J. B \textbf{16}, 85 (2000).   

\bibitem{Sei01} G. Seibold and J. Lorenzana, 
                    Phys. Rev. Lett. \textbf{86}, 2605 (2001).
                    
\bibitem{Bun13} J. B\"unemann, M. Capone, J. Lorenzana, and G. Seibold,
                     New J. Phys. \textbf{15}, 053050 (2013).                    
\bibitem{Jia12} C.~J. Jia, C.-C. Chen, A.~P. Sorini, B. Moritz, and T.~P. Devereaux,
                     New J. Phys. \textbf{14}, 113038 (2012).
                  
\bibitem{Wan14} in the Supplemental Material of
                Y. Wang, C.~J. Jia, B. Moritz, and T.~P. Devereaux,
                     Phys. Rev. Lett. \textbf{112}, 156402 (2014).


\bibitem{Che94} Y.~C. Chen, A. Moreo, F. Ortolani, E. Dagotto, and T.~K. Lee,
                     Phys. Rev. B \textbf{50}, 655 (1994). 

\bibitem{Buh99} C. Buhler and A. Moreo,
                     Phys. Rev. B \textbf{59}, 9882 (1999).
                     
\bibitem{Bec00} F. Becca, M. Capone, and S. Sorella,
                     Phys. Rev. B \textbf{62}, 12700 (2000).


\bibitem{Pre97} R. Preuss, W. Hanke, C. Gr\"ober, and H.~G. Evertz,
                     Phys. Rev. Lett. \textbf{79}, 1122 (1997).
                     
\bibitem{Gro00} C. Gr\"ober, R. Eder, and W. Hanke,
                     Phys. Rev. B \textbf{62}, 4336 (2000).
                     
\bibitem{Koh04} M. Kohno, X. Hu, and M. Tachiki,
                     Physica C \textbf{412-414}, 82 (2004).
                

\bibitem{Kun15} Y.~F. Kung, E.~A. Nowadnick, C.~J. Jia, S. Johnston, B. Moritz,
                R.~T. Scalettar, and T.~P. Devereaux,
                     Phys. Rev. B \textbf{92}, 195108 (2015).   
           

\bibitem{Ras88} J.~W. Rasul and T. Li,
                    J. Phys. C: Solid State Phys. \textbf{21}, 5119 (1988).
                    
           
\end{thebibliography}
\end{document}